# Recent advances on the spherical metal oxides for sustainable degradation of antibiotics

Ke Zhu[a,#], Xin Li[a,#], Yuwen Chen[a], Yizhe Huang[a], Zhiyu Yang[a], Guoqing Guan[c,*], Kai Yan[a,b,*]

[a] Guangdong Provincial Key Laboratory of Environmental Pollution Control and Remediation Technology, School of Environmental Science and Engineering, Sun Yat-sen University, Guangzhou 510275, China

[b] Guangdong Laboratory for Lingnan Modern Agriculture, South China Agricultural University, Guangzhou 510642, China

[c] Energy Conversion Engineering Laboratory Institute of Regional Innovation (IRI), Hirosaki University, 3-Bunkyocho, Hirosaki 036-8561, Japan

* Corresponding authors. *Email addresses*: guan@hirosaki-u.ac.jp (G. Guan), yank9@mail.sysu.edu.cn (K. Yan).

[#] Theses authors contribute equally to this work

**Abstract:**

Due to the permanent harm to human health and ecosystem balance, antibiotic pollution in water has become an important direction of current environmental governance. Spherical metal oxides (SMOs) have been frequently utilized as effective heterogeneous photocatalysts for the efficient degradation of antibiotics due to the unique properties (e.g., strong light absorption ability, high separation efficiency of photo-generated electron hole pairs, and good catalytic activity). This review will firstly focus on summarizing the rational design and synthesis of SMOs with various tuned microstructures such as hollow, porous shell, yolk shell, core shell, and nanoflowers. These structures can expose more active sites, achieve a higher utilization rate of light, enhance the mass transfer efficiency and improve the effective diffusion of reactive oxygen species (ROS). Secondly, this review will mainly analyze the intrinsic relationship between the structure of SMOs and its photocatalytic property, the ability to generate ROS, and the degradation pathway for antibiotics. Moreover, the photocatalytic mechanisms and recent progress of different SMOs catalysts for degrading typical antibiotics are compared in detail. Finally, challenges and prospects of future direction in the development of SMOs for antibiotic degradation are reviewed. It is expected to provide a rational design of SMOs catalysts for efficient photocatalytic degradation of environmental pollutants.

**Contents**

## 1. Introduction

With the social development, antibiotics are frequently used in people's daily life [1, 2]. According to the report of World Health Organization (WHO), more than 14,272 tons of antibiotics were consumed in 2018 based on the data mainly collected from imports, reimbursements, and sales in five regions [3]. Generally, antibiotics are often utilized in pharmaceutical and personal care product manufacturing, livestock industry, agriculture, fishing, living, medical solid and liquid waste systems, etc. [4] **(Fig. 1)**. Albeit antibiotic concentrations in the environment are usually low, they cannot be degraded naturally and chronically exist in our environment, with the potential harm for chronic toxicity to organisms in the ecological environment. To date, a total of 145 quantitative records of antibiotics and antibiotic resistance genes (ARGs) in coastal and estuarine environments have been collected (**Fig. 2a**). How to remove these antibiotics is of crucial importance and has attracted a lot of attention in the world. **Fig. 2b** shows the number of studied antibiotics in various environmental zones and basins in China (2009-2019). These antibiotics widely exist in different areas of China and seriously influence water safety.

At present, the conventional antibiotic removal processes mainly include physical [5, 6], biological [7, 8], and ecological and chemical methods [9, 10]. Among them, advanced oxidation processes (AOPs) are a more effective way to deal with antibiotic contamination. Usually, AOPs are featured by the generation of reactive oxygen species (ROS) with strong oxidizing abilities, by which macromolecular stubborn organic compounds can be effectively oxidized to low- or non-toxic small molecules under different reaction conditions (temperature, pressure, electricity, light, sound, catalysis) [11, 12]. Recently, photocatalysis mediated by heterogeneous semiconductors has been considered an efficient and low-cost method for solving environmental pollution and energy crisis. However, the rational design of a proper photocatalyst for the heterogeneous degradation process is still crucial.

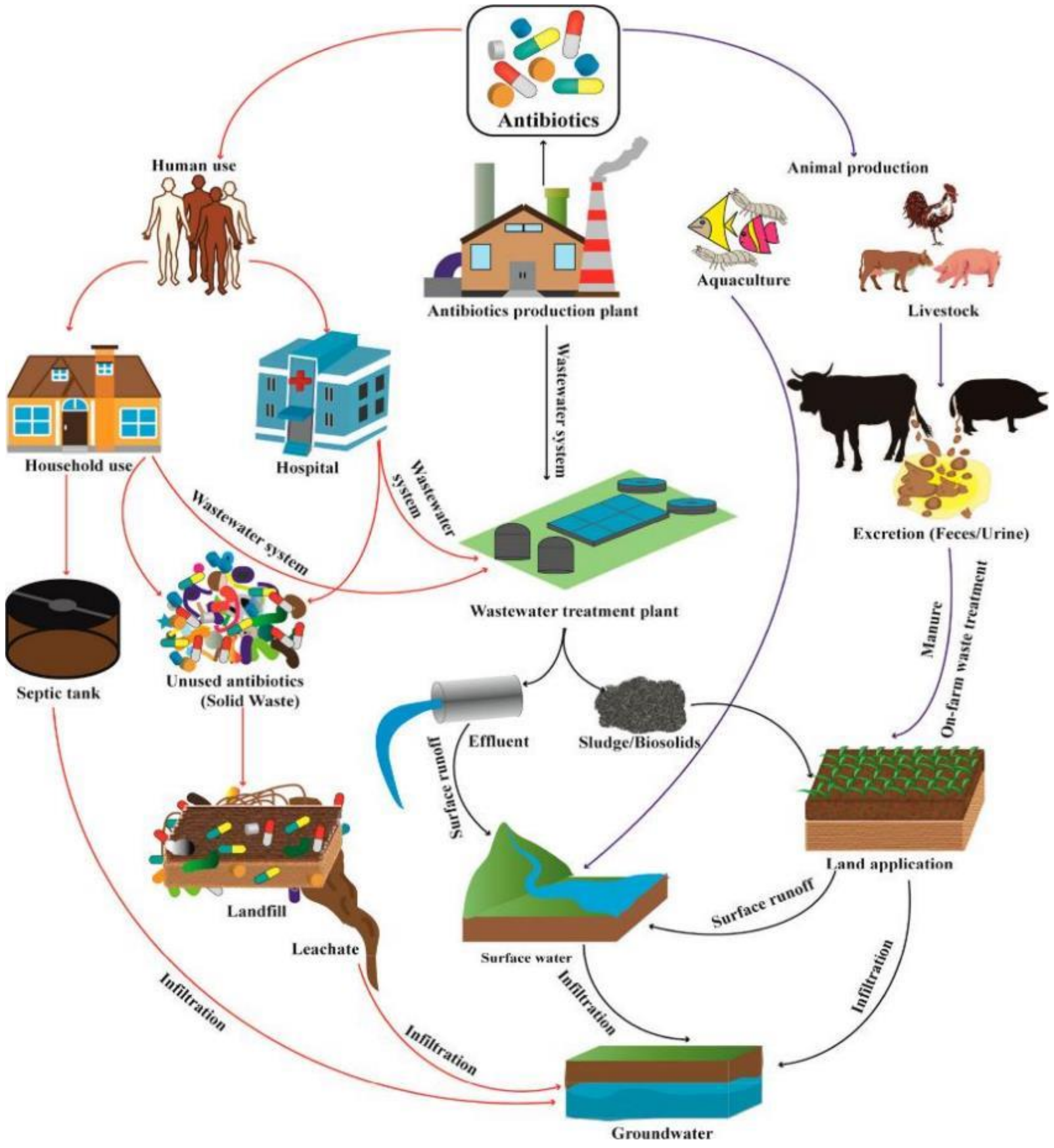

**Fig. 1.** The main sources and pathways of antibiotics occurrence in the aquatic environment. Reprinted with permission from Ref. [4].

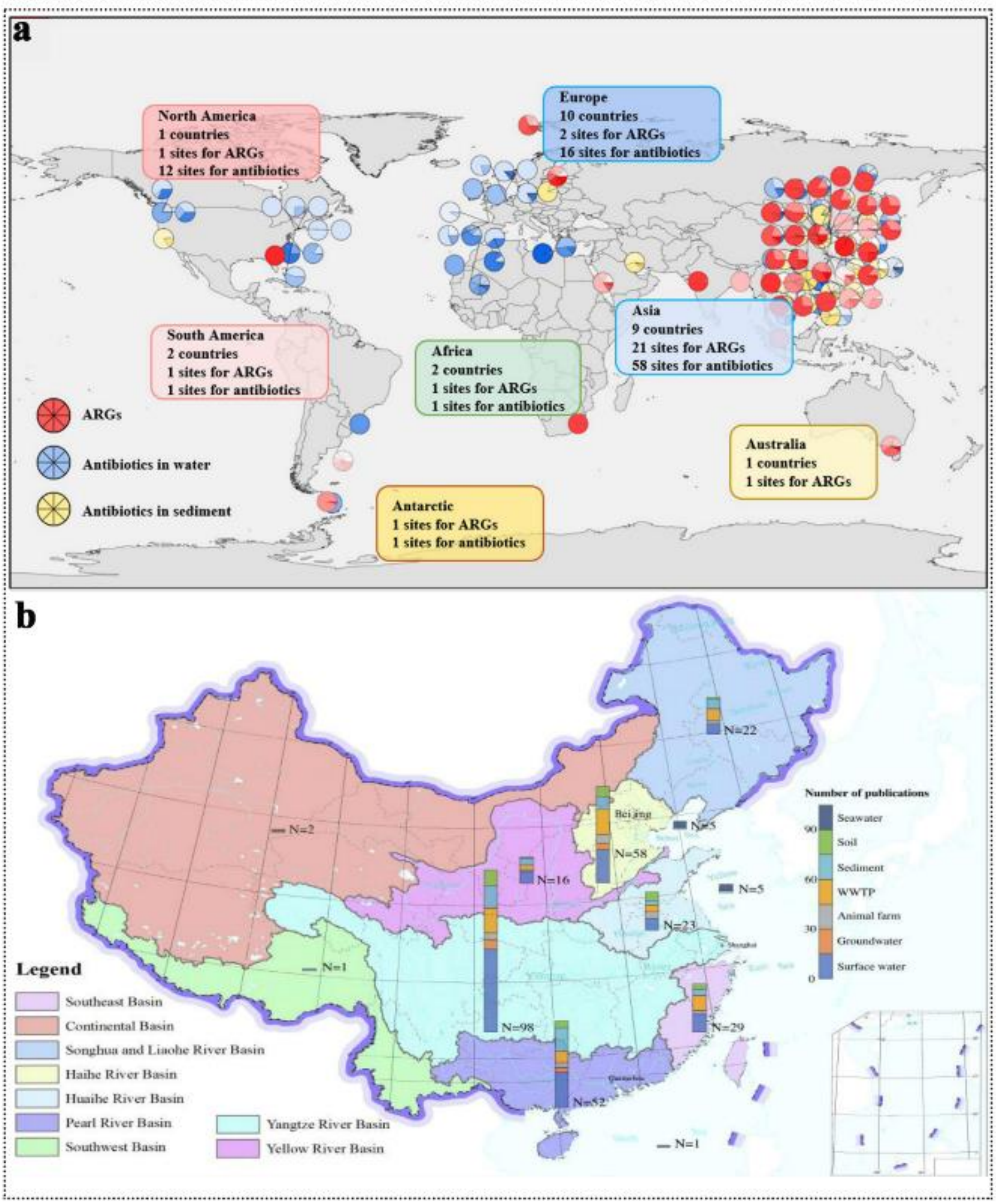

**Fig. 2.** (a) Geographical distributions of antibiotics and ARGs in estuarine and coastal environments of the world. Reprinted with permission from Ref. [13]. (b) The number of antibiotics in various environmental zones and basins in China from 2009 to 2019. Reprinted with permission from Ref. [14].

Over the last few decades, various inorganic semiconductor materials have been successively developed. According to the composition, they can be divided into oxides ($TiO_2$ [15, 16], $WO_3$ [17], $Co_3O_4$ [18], etc.), nitrides (g-$C_3N_4$ [19, 20], carbon nitride foam [21], etc.) and sulfides (CdS [22, 23], ZnS-CdS [24], etc.) and inorganic acid salts ($NiCo_2O_4$ [25, 26], $Bi_3NbO_7$ [27], perovskite [28], spinel [29, 30], etc.). Among them, metal oxides (MOs) semiconductors have been regarded as effective photocatalysts for antibiotic wastewater purification with more efficient utilization of active centers, acceleration of electron transportation, and regulation of defects [31]. Besides, MOs with three-dimensional (3D) spherical structures (micro flower, core-shell structure, and hollow sphere) have attracted much attention as photocatalysts to degrade antibiotics because of their advantages as follows: (1) Abundant active sites. In comparison with traditional one-dimensional and two-dimensional semiconductor materials, 3D spherical metal oxides (SMOs) usually have larger surface areas and tuned porous structures, which can provide many adsorption and reaction active sites. In particular, adsorption can realize the enrichment of pollutants in water, make the oxidation reaction proceed on the catalyst surface, and improve the mass transfer efficiency of ROS. (2) High utilization rate of light. The hollow cavity inside the SMOs with a hollow structure enables multiple scattering and reflection of light, which will improve the utilization rate of light. (3) Stability. The 3D sphere structure has good physical, chemical, and thermal stability. It can effectively reduce photo-corrosion to improve catalysis stability. (4) High quantization efficiency. Mixed/composite metal oxide spheres can form a heterojunction interface between different MOs, improving their quantization efficiency and light response range. (5) High production of radicals. Semiconductor-mediated SMOs photocatalysts enable good separation of reductants ($e^-$) in the conduction band (CB) and oxidants ($h^+$) in the valence band (VB) under light irradiation, resulting in releasing various ROS [32-34].

As well known, the structure of catalysts plays one of vital roles in the removal of antibiotics and how to especially synthesize the spherical structure is crucial. To date, the synthesis strategies can be generally divided into four groups including hard-, soft-, self-template [35-38], and template-free methods [39, 40]. Generally, the synthesis methods of SMOs are optimized from three aspects to improve their catalytic properties. (1) Well-controlled size and shape: spherical micro/nanostructured and diverse internal geometries (hollow, porous, mesoporous, delaminated, etc.) [41-43]. (2) Well-tuned material properties: specific surface area, channel structure, high porosity, low density, and shell permeability. (3) Well-designed surface engineering: the energy band structure, surface modification, edge defects, crystallinity, crystal structure, oxygen vacancy, etc. [15, 22, 44-46]. Because of their unusual structure, SMOs have received significant growth and wide applications in electrocatalysis, chemical sensors, photocatalysis, water splitting, and lithium-ion batteries [39, 47-49]. Although it is not possible to include every article exhaustively, an overall upward trend in SMOs published research has been observed (**Fig. 3**). So far, previous reviews have been mostly focused on the preparation, modification, and application of SMOs [39, 47-49]. There are few critical reviews on the structure-activity relationship between the photocatalytic mechanism and morphologies structure of SMOs for the degradation of antibiotics

From this perspective, this review will focus on the latest progress in the synthesis of SMOs catalysts and their utilization in the efficient photocatalytic degradation of antibiotics. Firstly, the synthesis methods of SMOs for tuning different microstructures are summarized. The synthetic mechanism, physical, chemical, optical properties, and applications of SMOs are then comprehensively discussed. Subsequently, the photocatalytic performance and mechanisms of different types of SMOs (porous spheres, micro flowers, yolk-double shell spheres, yolk-shell spheres, hollow spheres, and hollow multi-shelled spheres) for the degradation of antibiotics (tetracyclines, quinolones, sulfonamides, and other antibiotics) are outlined. Ultimately, the current challenges and future perspectives on SMOs catalysts for water purification are also proposed, and it is expected to provide new ideas for further research.

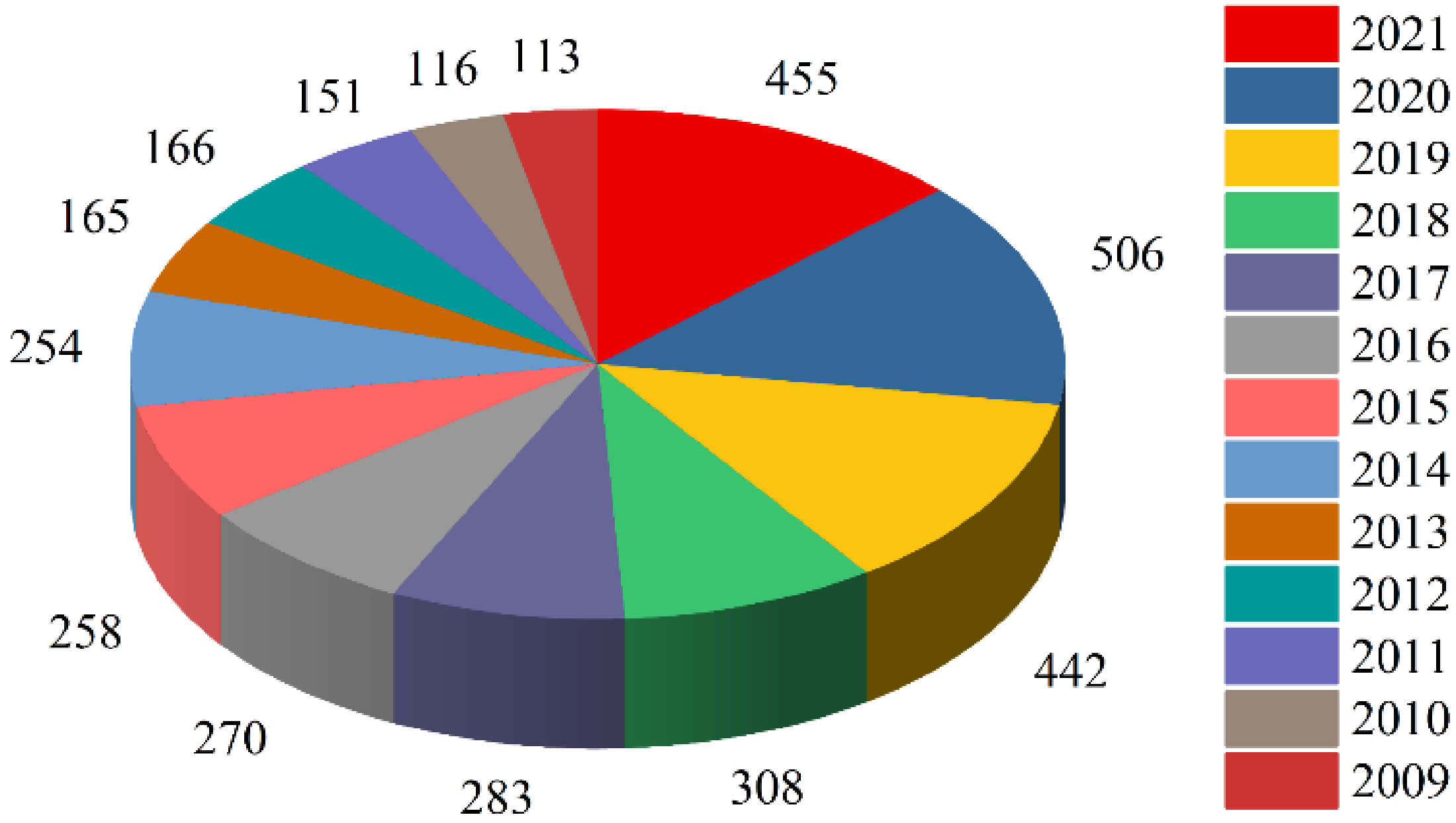

**Fig. 3.** Number of publications about the SMOs found in the Scopus database from 2009 to 2021. Reprinted with permission from Refs. Scopus database.

## 2. Synthesis strategies of SMOs

### *2.1. Template method*

The templating method has been widely used in synthesizing SMOs with different morphologies and structures, in which the morphology of product is mainly changed by controlling the nucleation and growth of crystals during the synthesis. Generally, the route of SMOs using the template method is divided into template synthesis, dispersion of metal precursors, and template removal. The features of templates can be generally classified into the hard-template, soft-template, and self-template methods.

#### *2.1.1. Hard-template method*

The hard-template method, also referred to as "nano casting", has been widely applied in nanomaterials synthesis since the early 1990s. Due to the tunable structure morphologies and sizes of hard templates, the synthesis of SMOs can achieve great regularity and desired structure, and in particular, structural collapse can be avoided as hard templates have relatively high rigidity. Typical hard templates for spherical materials preparation include $SiO_2$, carbonaceous spheres, MOs, polymers, etc. Metal precursors are attached to the template surface via electrostatic force, chemical bonding, or van der Waals force, followed by particle assembling and growth to form the special structure of SMOs, which naturally inherit the replica of spherical templates. Then, templates are removed with corresponding methods based on the physicochemical property of the templates, among which chemical etching (with NaOH, HF, or $NH_4HF_2$) and thermal treatment are widely applicable (**Fig. 4**).

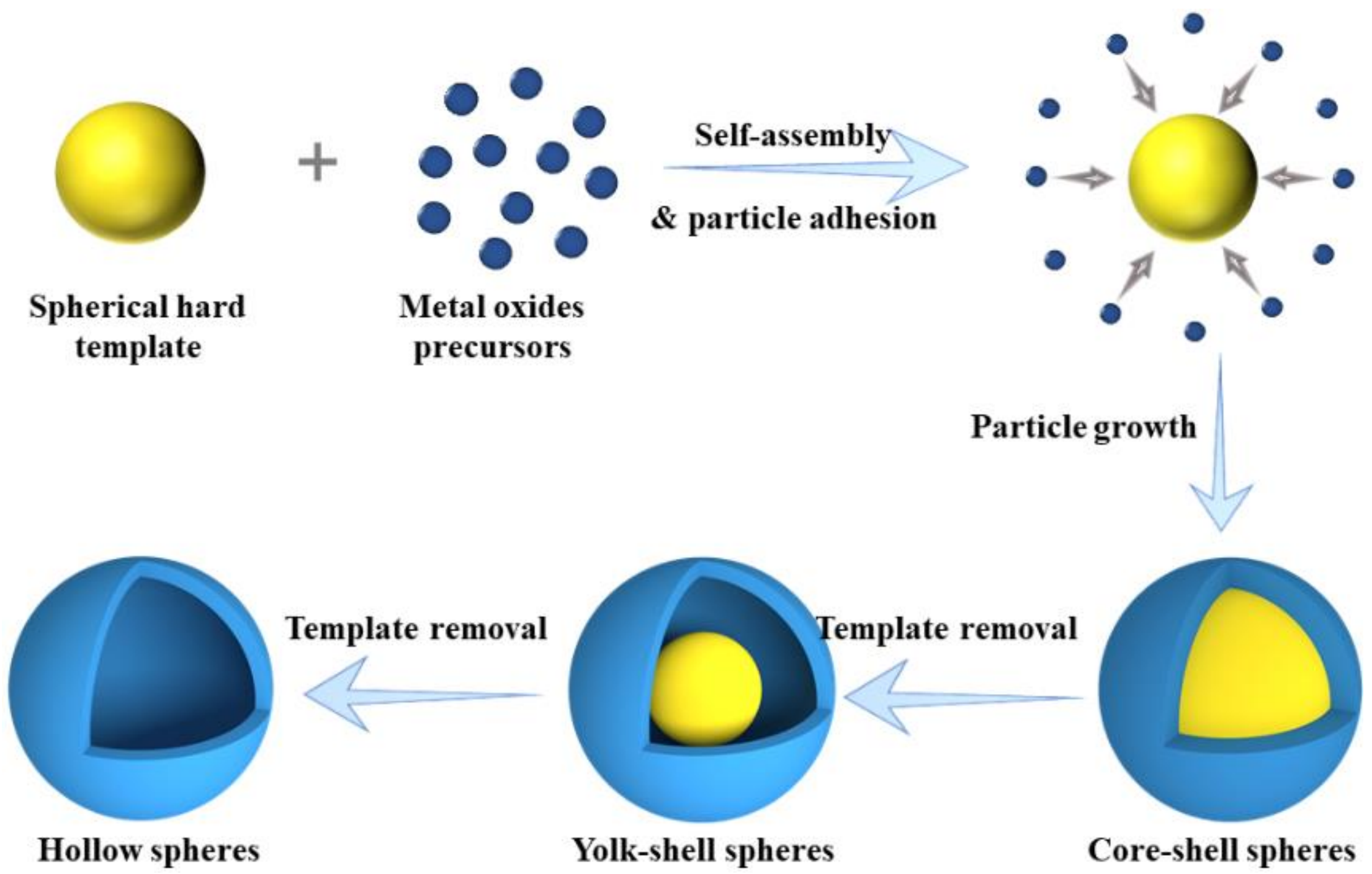

**Fig. 4.** Diagram for the synthesis of hollow SMOs by a hard-template method. Reprinted with permission from Ref. [50].

The hard-template method has the advantages of tuning size, alternatively, changing the sphere size and shell thickness. Lee et al. [51] synthesized single-atom Cu/$TiO_2$ hollow spheres with $SiO_2$ as a hard template (**Fig. 5a**). In the process, Cu/$TiO_2$ spheres were coated with tetraethyl orthosilicate (TEOS) to prevent surface diffusion, which stabilized the Cu atom on the $TiO_2$ surface. Kanjana et al. [52] synthetized hollow mesoporous $TiO_2$ spheres (THs) with carbon sphere (CS) as the hard template (**Fig. 5b**). Adjusting the initial pH of the solution from 3 to 10, spherical size can be flexibly controlled ranging from 171 to 668 nm. As for photocatalytic degradation of antibiotics, a hollow structure could provide a high specific surface area, which exposes more reactive sites and facilitates pollution diffusion. On the other hand, such a structure allows light reflection within spheres and benefits faster electron transfer on the surface, leading to efficient light absorption and electron-hole separation. Wang et al. [53]

synthesized BiOCl@$CeO_2$ hollow microsphere through the precipitation-hydrothermal process also using the carbon sphere as the hard template (**Fig. 5c**). When it was used for degrading tetracyclines with visible light, the removal rate within 120 min was encouragingly up to 92%.

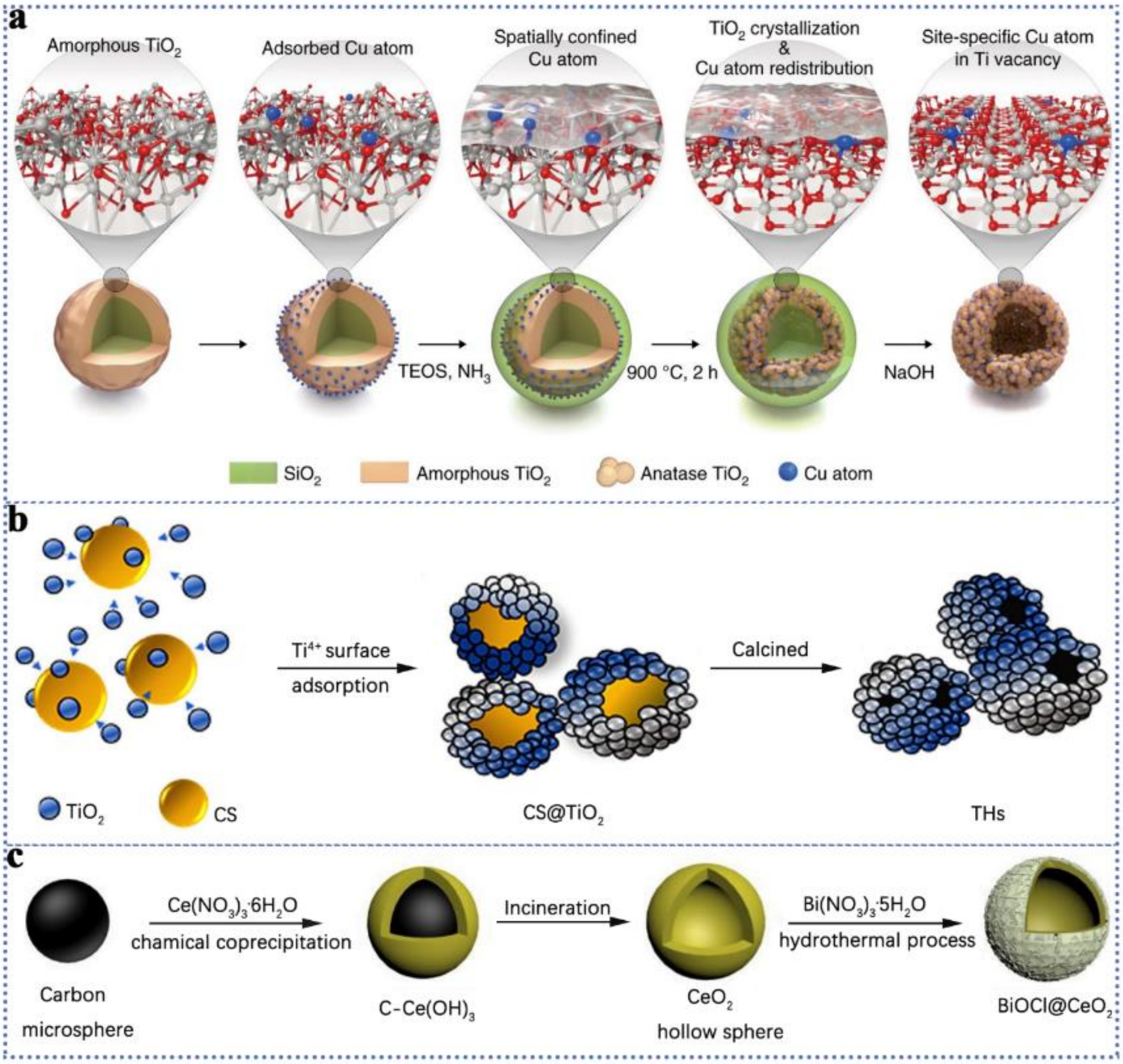

**Fig. 5.** Typical examples of synthesizing SMOs by a hard-template method: (a) Synthetic process of Cu/$TiO_2$ hollow spheres. Reprinted with permission from Ref. [51]. (b) $TiO_2$ hollow mesoporous sphere. Reprinted with permission from Ref. [52]. (c) BiOCl@$CeO_2$ hollow spheres. Reprinted with permission from Ref. [53].

In recent years, more researchers have focused on the development of multi-shell hollow SMOs. Compared with single-shell hollow spheres, hollow multi-shelled structures (HoMSs) can more efficiently promote light absorption by allowing multiple light reflections between adjacent shells, and the specific surface area is also largely enhanced by the unique structure. Wang et al. [54, 55] proposed a concept of a sequential template approach (STA) based on the hard templating method for manufacturing HoMSs, which is facile and has universality for various SMOs syntheses. The general process of STA preparing HoMSs includes two steps: (1) metal ions precursors are adsorbed to the hard template surface, and partially diffused into the hard template under the effect of a concentration gradient; and (2) template is then removed by the gradient heating calcination [56]. Zhang et al. [57] successfully fabricated $WO_3$ HoMSs with up to three shells (**Fig. 6a**), whereas shell thickness was limited to 35-90 nm by optimizing the distribution of W ions adsorbed in the CS.

Apart from the conventional non-metallic templates, SMOs can also be used as hard templates for the in-situ synthesis of multi-shell SMOs. For example, Wei et al. [58] constructed $SrTiO_3$-$TiO_2$ as a photocatalyst. **Fig. 6b** shows that $TiO_2$ HoMSs served as a hard template. $Sr^{2+}$ can infiltrate the $TiO_2$ sphere and adhere to the surface in an alkaline solution, then react with $TiO_2$ to form perovskite $SrTiO_3$ through a hydrothermal treatment. The advantages of this method include: (i) the removal of hard templates can be eliminated; (ii) the heterojunction could be constructed with $SrTiO_3$, which can establish additional paths for photocarriers and holes with the improved separation efficiency of charge transfer and photocarriers, thereby enhancing the performance.

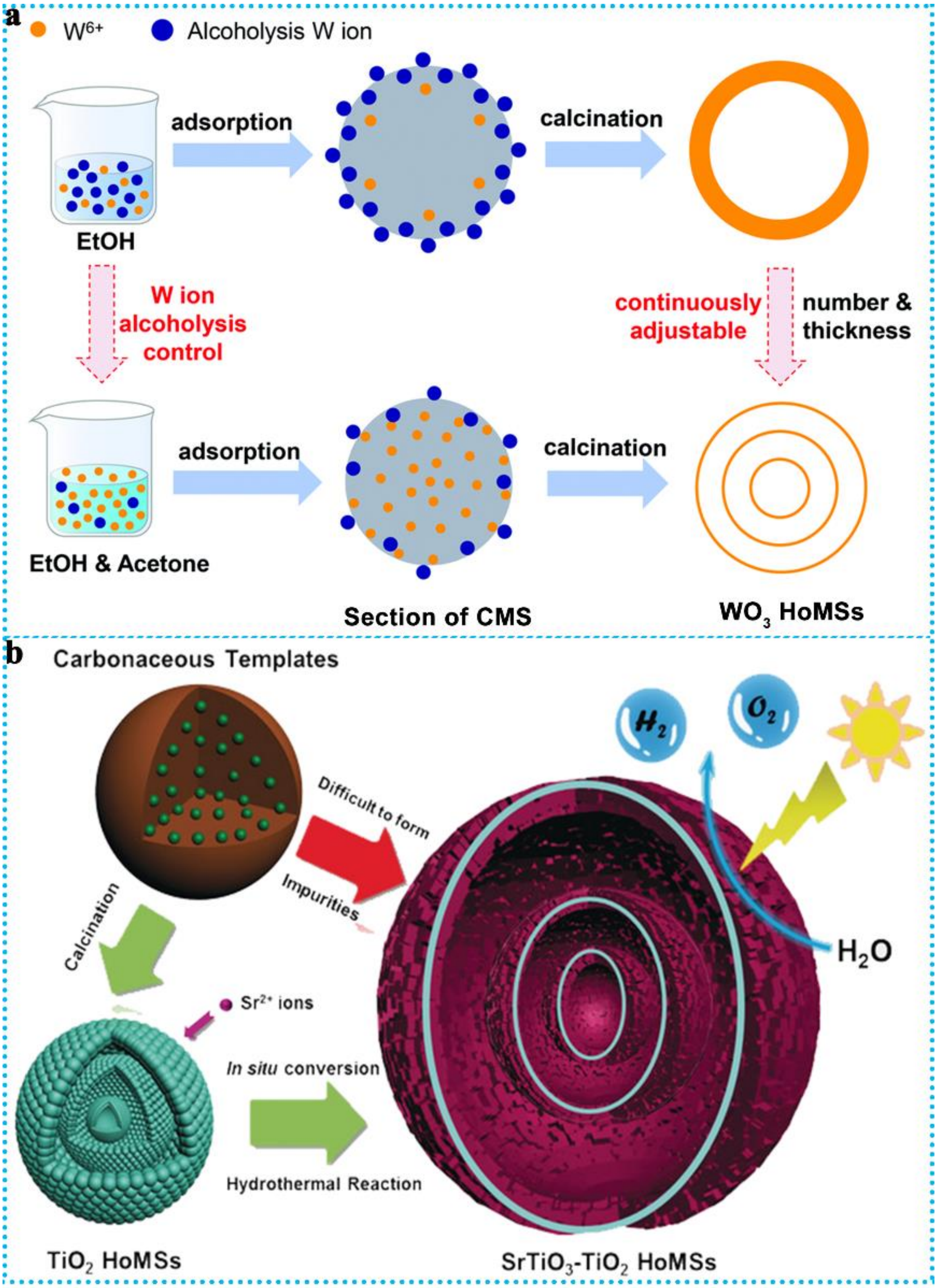

**Fig. 6.** The synthesis of HoMSs by a hard-template method: (a) $WO_3$ with CS as hard templates. Reprinted with permission from Ref. [57]. (b) $SrTiO_3$-$TiO_2$ with $TiO_2$ as hard templates. Reprinted with permission from Ref. [58].

**Table 1** The synthesis of SMOs with different hard templates.

| Contents | Template | Precursors | Synthetic condition | Morphology | Refs. |
|---|---|---|---|---|---|
| $TiO_2$ | CSs | TBT | • 10 g $L^{-1}$ CSs mixed with 6.694 mmol TBT in 20 mL ethanol<br>• Aging for 18 h<br>• Calcined at 450 °C for 1 h | 1. Hollow mesoporous sphere<br>2. AD: 668-171 nm<br>3. SSA: 138-54 $m^2$ $g^{-1}$ | [52] |
| BiOCl @$CeO_2$ | CSs | $Ce(NO_3)_3·6H_2O$, $NH_4Ac•2H_2O$ | • 4 g $L^{-1}$ CSs mixed with precursors<br>• Heated at 60 °C for 12 h<br>• Calcined at 450 °C for 30 min | 1. Hollow sphere<br>2. AD: 200-300 nm<br>3. SSA: 53.4 $m^2$ $g^{-1}$ | [53] |
| $WO_3$ | CSs | $WCl_6$ | • 0.6 g CSs mixed with 0.6 g $WCl_6$ in ethanol<br>• Followed by fabrication with an STA | 1. Hollow multi-shelled structures<br>2. Shell thickness: 35-90 nm<br>3. SSA: 41.0-62 $m^2$ $g^{-1}$ | [57] |
| CT-g-ZnTAPc | $SiO_2$ | $Cu(NO_3)_2$, TBOT, MAH | • Remove $SiO_2$ with 1 M NaOH solution | 1. Hollow sphere<br>2. AD: 260-330 nm<br>3. SSA: 219.3 $m^2$ $g^{-1}$ | [59] |
| Cu/$TiO_2$ | $SiO_2$ | TBOT, $CuCl_2•2H_2O$ | • Modified wrap-bake-peel process | 1. Hollow sphere<br>2. AD: ~250 nm | [51] |
| Pt/$TiO_2$ -$ZrO_2$ | Polystyrene (PS) | $C_{16}H_{36}O_4Zr$, $H_2PtCl_6$, $C_{12}H_{28}O_4Ti$ | • Two-step vacuum impregnation method<br>• Calcined at 500 °C for 7 h | 1. Hollow sphere<br>2. AD: 400 nm<br>3. SSA: 110.7 $m^2$ $g^{-1}$ | [60] |
| $SiO_2$-$Fe_2O_3$ @$TiO_2$ | Cationic PS (CPS) | $Fe(NO_3)_3$, TBT | • Remove CPS by pyrolysis at 450 °C | 1. Hierarchical hollow sphere<br>2. AD: ~500 nm<br>3. SSA: ~180 $m^2$ $g^{-1}$ | [61] |
| $SrTiO_3$ -$TiO_2$ | $TiO_2$ | $Sr(OH)_2•8H_2O$ | • 79 mg $TiO_2$ immersed with 1mmol precursors<br>• Heated at 160 °C for 4 h | 1. Hollow multi-shelled structures<br>2. AD: 45 nm | [58] |

All these results show that the hard template method is beneficial to the preparation of hollow SMOs. It is well-known that SMOs with a hollow structure enable multiple scattering and reflection of light, which can improve the utilization rate of light. **Table 1** shows recent developments in the synthesis of SMOs with different hard templates. Albeit the hard-template method has achieved great progress in obtaining uniform, controllable, and highly crystalline SMOs. It still faces the following drawbacks: (1) hard template synthesis is relatively complicated and requires much effort, which is not conducive to scale production; (2) template removal can be intractable since it usually involves the use of corrosive acid or alkali; and (3) it is difficult to find templates with the required size, shape, and surface characteristics when growing hollow spheres from target materials.

#### *2.1.2. Soft-template method*

The soft-template method makes use of macromolecular aggregate which can spontaneously assemble into specific structures, acting as templates in the synthesis process. The soft-template method has more advantages in template removal and flexible morphology control than the hard-template method. Soft templates for SMOs synthesis include surfactants, block copolymers, organic molecules, and biological templates (**Fig. 7a**). These templates are the clusters with a certain space limit, controlling the self-assembly of metal precursors in a certain order by their intermolecular force and spatial limiting ability, to achieve the controllable synthesis of material morphology, structure, size, and arrangement (**Fig. 7b**).

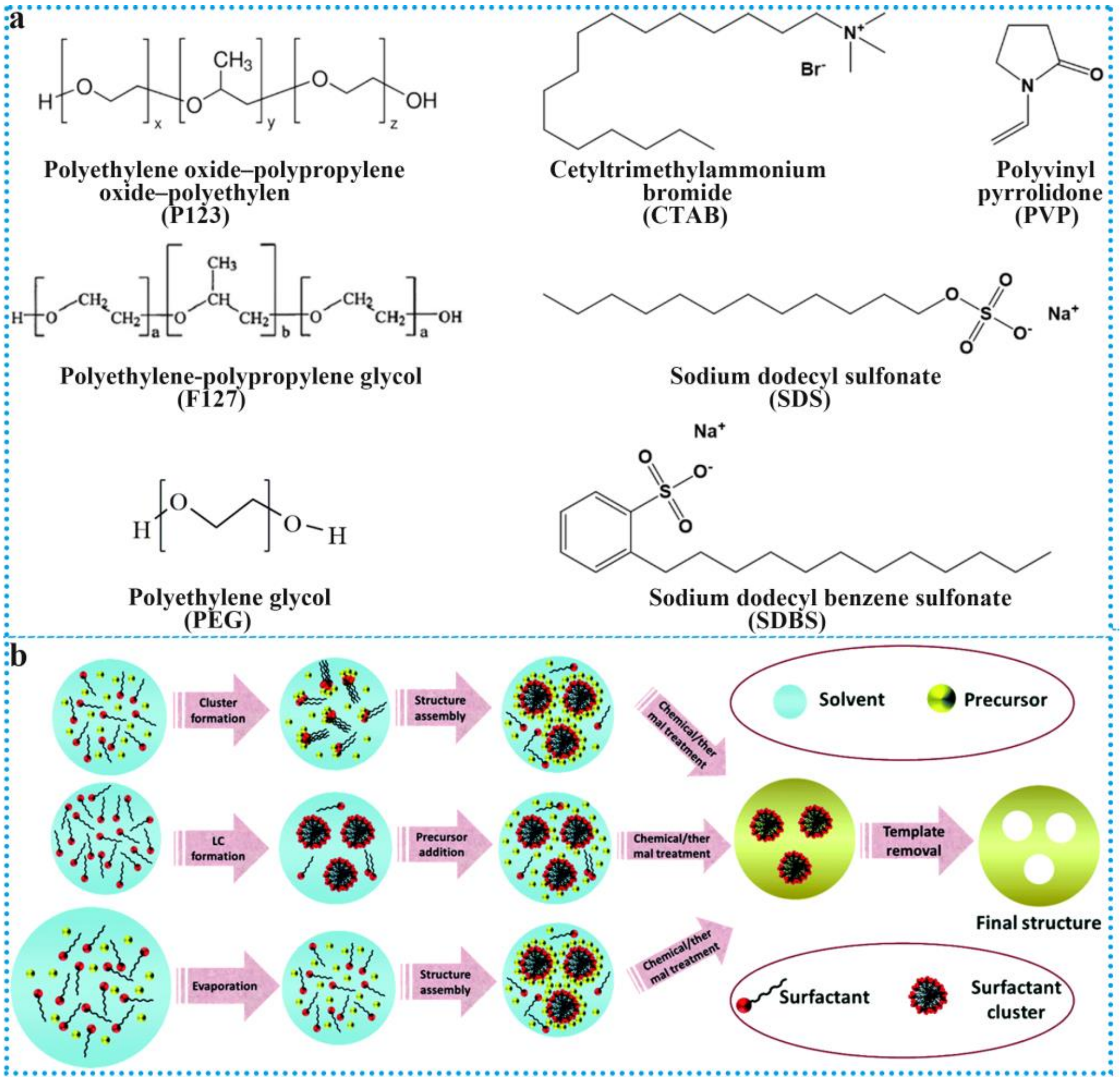

**Fig. 7.** (a) Common soft template formula. Reprinted with permission from Ref. [62-68]. (b) The general mechanism for soft-template method. Reprinted with permission from Ref. [69].

Surfactants are amphiphilic compounds containing both hydrophobic and hydrophilic groups, which significantly decrease the interfacial tension. They can self-assemble to form micelles or vesicles, and metal precursors are adsorbed on their surface through electrostatic adsorption or chemical adsorption, constructing a single or multi-layer shell. As a typical surfactant, cetyltrimethylammonium bromide (CTAB) is ideally considered for microsphere construction. For example, Zhong et al. [70] successfully synthesized $Cu_2O$/Cu@CoO hierarchical nanosphere with CTAB as the soft template. In such a process, as the CTAB concentration surpassed its critical micelle concentration (CMC) within a certain range, the template molecules aggregated dynamically, forming spherical micelles. $Cu^{2+}$ and $Co^{2+}$ were guided to aggregate around the interface between a hydrophilic group of CTAB and the solution, successively copying the spherical structure of micelles, then by calcination and Ostwald ripening, the $Cu_2O/Cu@CoO$ nanosphere with core-shell structure was obtained. Zou et al. [71] also used CTAB to prepare ZnO flower-like microspheres for photocatalytic degradation of organic dye (**Fig. 8a**). The constructed ZnO had a thin nanosheet with a thickness of around 30 nm, which greatly enlarged the specific surface area.

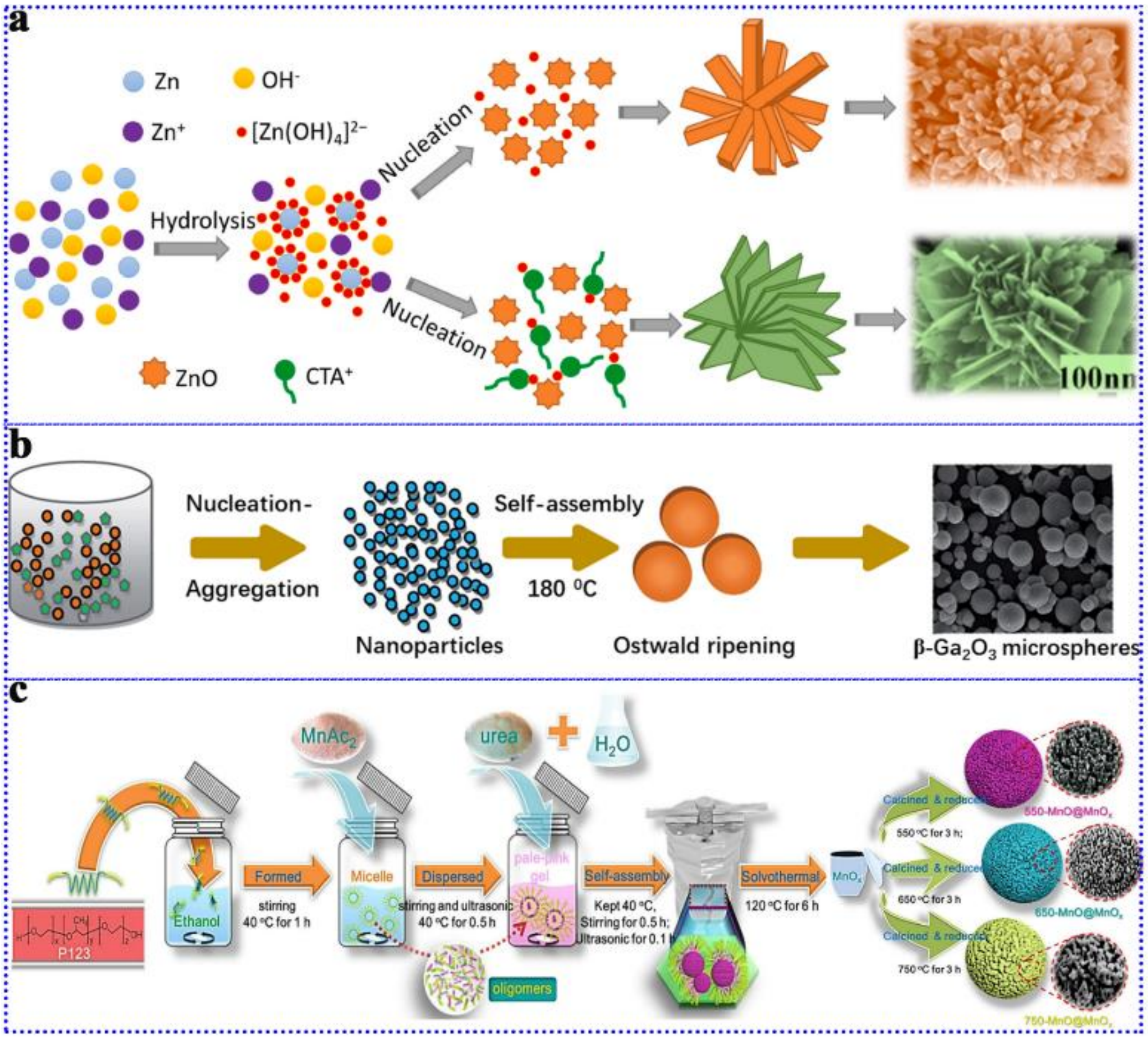

**Fig. 8.** SMOs synthesized with the soft-template methods: (a) ZnO hierarchical microflowers. Reprinted with permission from Ref. [71]. (b) β-$Ga_2O_3$ microspheres.

Reprinted with permission from Ref. [72]. (c) $MnO@MnO_x$ mesoporous microsphere. Reprinted with permission from Ref. [73].

Block copolymers consist of two or more polymer segments with different properties by covalent bonds, which are widely employed in the soft-template method. Typical block copolymers include polyethylene-polypropylene glycol (F127) and polyethylene oxide-polypropylene oxide-polyethylene (P123). Girija et al. [72] used F127 to synthesize single-crystalline photocatalyst $\beta$-$Ga_2O_3$ microspheres ranging from 1 to 3 μm for Rhodamine B degradation (**Fig. 8b**). P123 is an amphiphilic tri-block copolymer composed of poly(ethylene oxide) (PEO) and poly(propylene oxide) (PPO), which can flexibly form micelles or multi-layer vesicles in an aqueous solution by self-assembly, and hence multiple spherical structures can be synthesized by precisely controlling the behavior of P123. Our team has used P123 as the soft template to successfully obtain mesoporous $MnO@MnO_x$ microspheres (**Fig. 8c**). It is found that P123 surfactant can greatly enhance the string interaction of the P123 template with the metal ions atoms and molecules, leading to a more uniform and rapid heating process, and added urea can promote the esterification reaction during the solvothermal treatment. The obtained $MnO@MnO_x$ microspheres with porous structures consist of nanoparticles, which increases the specific surface area to 167.7 $m^2$ $g^{-1}$ and thus provides more active sites for photocatalytic degradation [73].

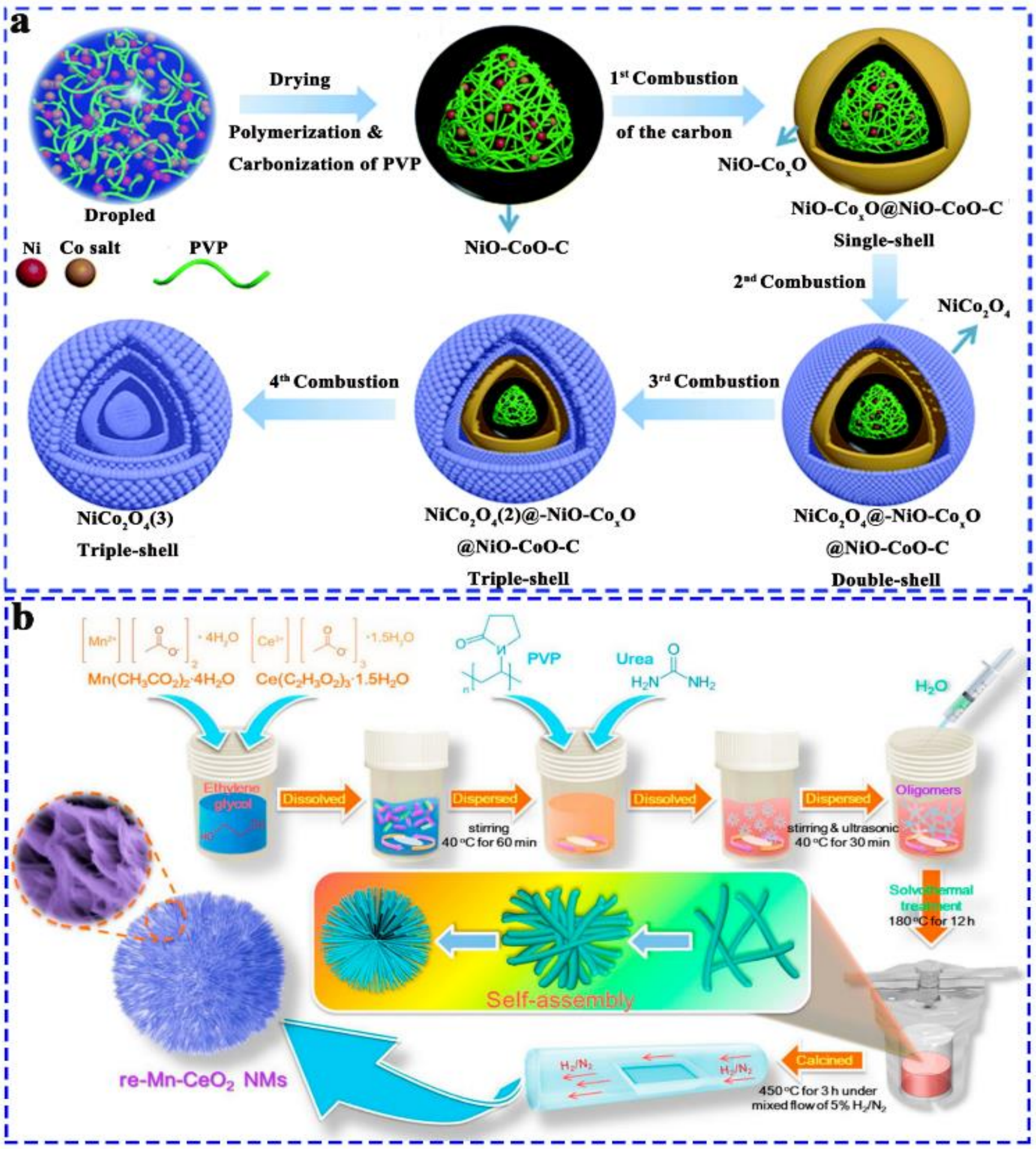

**Fig. 9.** (a) $NiCo_2O_4$ multishell yolk-shell spheres. Reprinted with permission from Ref. [74]. (b) Mn-$CeO_2$ hierarchical microflowers. Reprinted with permission from Ref. [75].

Polyvinyl pyrrolidone (PVP) is a kind of non-ionic water-soluble polymer, and it can form micelles with hydrophobic nuclei and hydrophilic shells in solution. Leng et al. [74] used PVP as the soft template to prepare $NiCo_2O_4$ spinel by a spray pyrolysis process (**Fig. 9a**). It is found that the amide group of PVP can complex some Ni/Co ions via strong ionic bonds. The subsequent polymerization results in the directional distribution of PVP and Ni/Co ions in the droplet layer by layer, then the multishell yolk-shell structure was produced by the graded pyrolysis and burning of PVP. Our team also reported that 3D hierarchical $H_2$-reduced Mn-doped $CeO_2$ microflowers (re-

$Mn-CeO_2$ NM) were assembled with nanotubes using PVP as the soft template (**Fig. 9b**), in which the staggering growth of nanotubes is conducive to the formation of a large number of small voids and openings, beneficial for the diffusion of organic pollutants [75]. Biological materials such as biomass, biomacromolecules, and some microbes can be also used as the soft template for nanomaterials synthesis. For example, Matos et al. [76] used biomass derivative furfural for controllable synthesis of $TiO_2$ anatase composed of microspheres with a mean size of ca. 4.0 ± 1.6 μm. Zhang et al. [77] took bacteria as the soft template and utilized its unique spatial structure and water and phosphorus-containing molecules to synthesize spheres composed of multi-layer $TiO_2$ nanosheets.

Overall, the soft-template method has been frequently used to generate SMOs with a variety of morphology and size because the template preparation and removal process is more facile by washing or calcination. **Table 2** shows the previously reported synthesis of SMOs with different soft templates. It can be seen that SMOs with structures including microspheres/nanosphere and microflowers were mainly generated by the soft-template method. The spheres as semiconductor materials have a 3D structure, which can enhance the specific surface area for exposing more adsorption and reaction active sites and enhancing the mass transfer efficiency of ROS and antibiotics. However, the soft-template method is sensitive to the condition of the preparation process, including stirring rate, ionic concentration, pH and solvent, etc. To precisely control the uniformity and morphology of SMOs, it should take more effort to investigate.

**Table 2** The synthesis of SMOs with different soft templates.

| Contents | Template | Precursors | Synthetic condition | Morphology | Refs. |
|---|---|---|---|---|---|
| $Cu_2O/Cu$ @CoO | CTAB | $CuSO_4•5H_2O$, $Co(NO_3)_2$ | • Mixed 0.1 M CTAB with precursors<br>• Precipitated with 0.2 M NaOH<br>• Calcined at 300 °C under $N_2$ atmosphere for 3 h | 1. Hierarchical nanosphere<br>2. AD: 10 nm | [70] |
| ZnO | CTAB | $Zn(NO_3)_2$ | • Hydrothermal treatment at 140 °C for 4 h | 1. Hierarchical microflower<br>2. AD: 1-4 μm | [71] |
| β-$Ga_2O_3$ | F127 | $Ga(NO_3)_3•nH_2O$ | • Hydrothermal treatment at 180 °C for 12 h<br>• Followed by calcined at 900 °C for 3 h | 1. Microsphere<br>2. AD: 1-3 μm | [72] |
| $MnO_x$ | P123 | $MnAc_2•4H_2O$ | • Hydrothermal treatment at 120 °C for 6 h<br>• Calcined at 450 °C for 6 h | 1.Mesoporous microsphere<br>2. AD: 1 μm<br>3. SSA: 89.2 $m^2$ $g^{-1}$ | [78] |
| MnO@ $MnO_x$ | P123 | $MnAc_2•4H_2O$ | • Mixed 0.1 M CTAB with precursors<br>• Hydrothermal reaction at 120 °C for 6 h<br>• Calcined at 450 °C for 3 h<br>• Reduced in $H_2$ for 10-30 min | 1. Mesoporous microsphere<br>2. AD: 2.45 μm<br>3. SSA: 167.7 $m^2$ $g^{-1}$ | [73] |
| Mn-doped $CeO_2$ | PVP | $MnAc_2•4H_2O$, $CeAc_3•1.5H_2O$ | • Hydrothermal reaction at 180 °C for 12 h<br>• Calcined at 450 °C for 3 h under $H_2/N_2$ atmosphere | 1. Hierarchical microflowers<br>2. AD: 2.5 μm<br>3. SSA: 430.3 $m^2$ $g^{-1}$ | [75] |
| $NiCo_2O_4$ | PVP | $Co(NO_3)_2•6H_2O$, $Ni(NO_3)_2•6H_2O$ | • One-pot spray pyrolysis | 1. Multishell yolk-shell structure<br>2. AD: 2.8 μm | [74] |
| C-$TiO_2$ | furfural | $C_{12}H_{28}O_4Ti$ | • Solvothermal treatment at 180 °C for 16 h<br>• Calcined at 550 °C for 5 h | 1. Microsphere<br>2. AD: 1.7 ± 0.9 μm<br>3. SSA: 13 $m^2$ $g^{-1}$ | [76] |
| $TiO_2$ | bacteria | TBT | Precipitation and calcination | 1. Nanosheet-assembled microsphere<br>2. AD: 0.6 μm<br>3. SSA: 33 $m^2$ $g^{-1}$ | [77] |

### *2.1.3. Self-template method*

In the past few years, exploration efforts have been put into self-template methods in developing SMOs. Compared with the hard/soft template methods, the self-template method is easier to remove the templates, thus saving the production cost, and effectively controlling the shell thickness and particle uniformity. Usually, In the process of constructing hollow nanostructures by the self-template method, the template material itself is involved in the formation of the shell, that is, the template material is directly transformed into the shell or as the precursor of the shell. Shen's groups [79] developed a self-template formation strategy for the fabrication of mixed-metal-oxide composite hollow structures. They used NiCo-glycerate spheres as the precursor by a simple nonequilibrium heat treatment process to prepare $NiCo_2O_4$ core-in-double-shell hollow spheres (**Fig. 10a**). Similarly, Ali et al. [80] also prepared $CuCo_2O_4$ microspheres with similar approaches, in which CuCo-glycerate precursor spheres were prepared via a facile solvothermal method and converted into $CuCo_2O_4$ spheres with different internal structures by adjusting the heating rate during calcination (**Fig. 10b**).

Liu et al. [81] reported that the formation of different hollow structures is mainly due to the different adhesion and contraction forces between interfacial layers in different calcination procedures. The obtained $ZnFe_2O_4$ hollow sphere material with a higher surface area and more effective light absorption possessed excellent photocatalytic performance for the degradation of gaseous o-dichlorobenzene. Likewise, Li et al. [82] successfully synthesized a novel 3D flower-like sphere $BiOBr/Bi_5O_7Br$ via the self-template method and studied the synthesis mechanism systematically. Ultimately, $BiOBr/Bi_5O_7Br$ showed excellent photocatalytic activity during antibiotic degradation due to its 3D stratification structure, optimal heterojunction ratio, and appropriate concentration of oxygen vacancy. Jiang et al. [83] first reported a flower-like $CuFe_2O_4$ nanostructure synthesized by the self-template method and used for the activation of peroxymonosulfate (PMS) to degrade carbamazepine. It was found that the flower-like $CuFe_2O_4$ was superior to other $CuFe_2O_4$ structures (nanoparticles, bulk, and sphere).

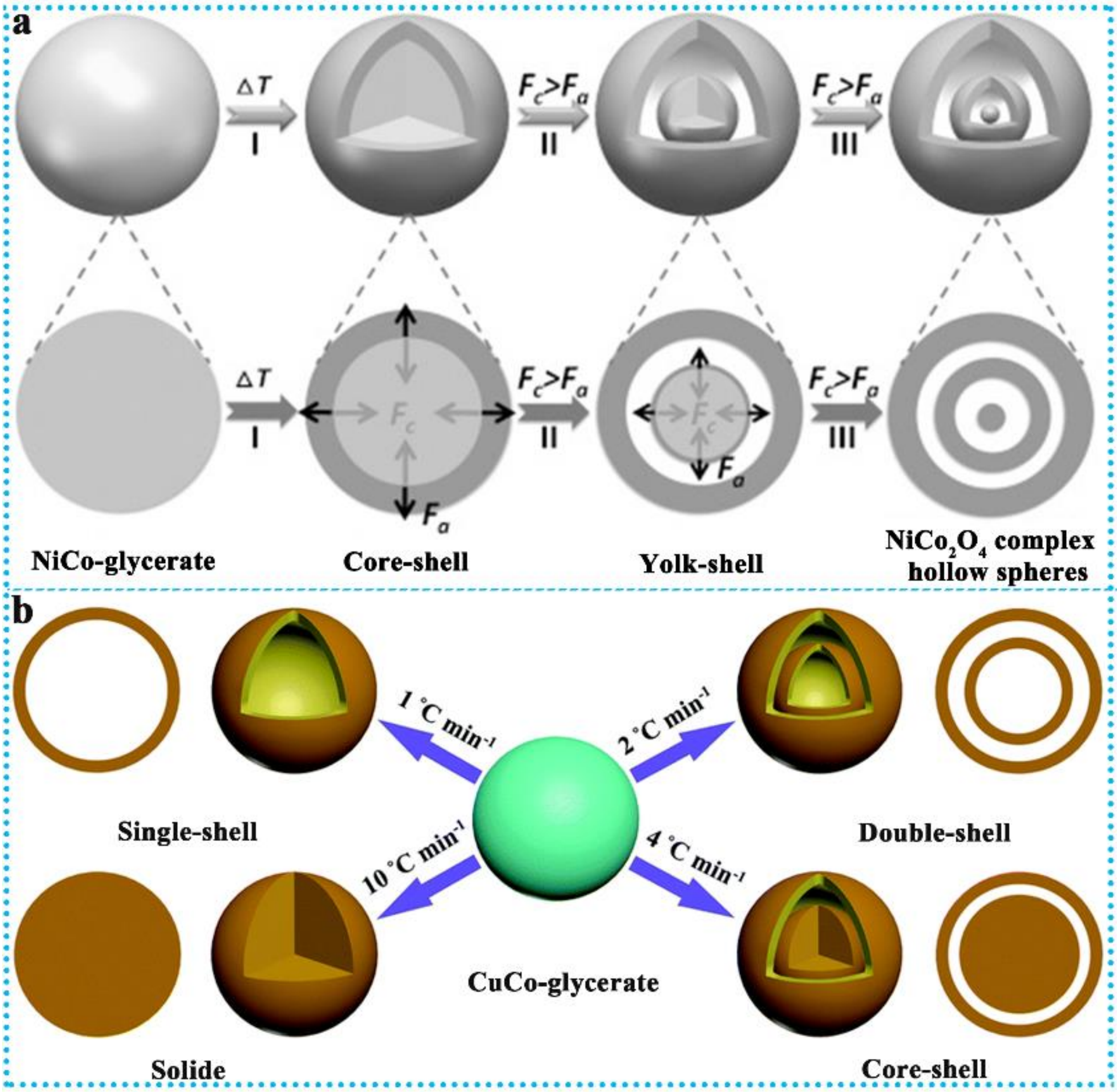

**Fig. 10.** The formation mechanism of SMOs with a self-template method: (a) $NiCo_2O_4$

core-in-double-shell hollow spheres. Reprinted with permission from Ref. [79]. (b) $CuCo_2O_4$ microspheres. Reprinted with permission from Ref. [80].

It is well-known that metal-organic frameworks (MOFs) are porous materials formed by the coordination of metal ions (as network nodes) and organic ligands (as linkers). MOFs have been widely applied in various applications due to their unique structure including outstanding porosity, high specific surface area, rich unsaturated metal sites, and low density [84-86]. Furthermore, the MOF derivatives prepared by organic cracking in the post-calcination template not only retain the unique structural characteristics of the precursor, but also produce SMOs with layered three-dimensional yolk-shell structure and hollow structure. Recently, MOFs have been proposed as an effective template for manufacturing SMOs. Therefore, Wang et al. [87] successfully synthesized a unique CuO@NiO multilayer hollow microspheres by the microwave thermal treatment of Cu-Ni bimetallic organic frameworks (**Fig. 11a)**. This microspheres size was ~0.9-1 μm and consisted of interconnected CuO/NiO nanoparticles and three layers spherical structure. Saleki et al. [88] chose a bimetallic copper-cobalt zeolitic imidazolate framework (CuCo-ZIF) as a templated of MOFs to prepare double-shell $CuCo_2O_4$ hollow spheres via an annealing treatment under air atmosphere (**Fig. 11b)**. The $CuCo_2O_4$ showed a large surface area of 93 $m^2$ $g^{-1}$. Meanwhile, Chen et al. [89] reported 3D yolk shell-like structure Ni@carbon was synthesized by pyrolyzing Ni-MOFs as a self-template, Inheriting the excellent performance of Ni-MOFs, including large surface area, rich Ni content, and porous layered yolk shell-like structure (**Fig. 11c**). Guo et al. [90] designed and developed $Co_3O_4$ microspheres using a simple C/Co-MOFs as self-template via two-step method including hydrothermal and calcination. Importantly, the size of C/Co-MOFs derived $Co_3O_4$ microspheres could be controlled by adjusting the content of PVP. When the added amount of PVP was 0.1 g, the $Co_3O_4$ sample showed a uniform microsphere structure (**Fig. 11d**).

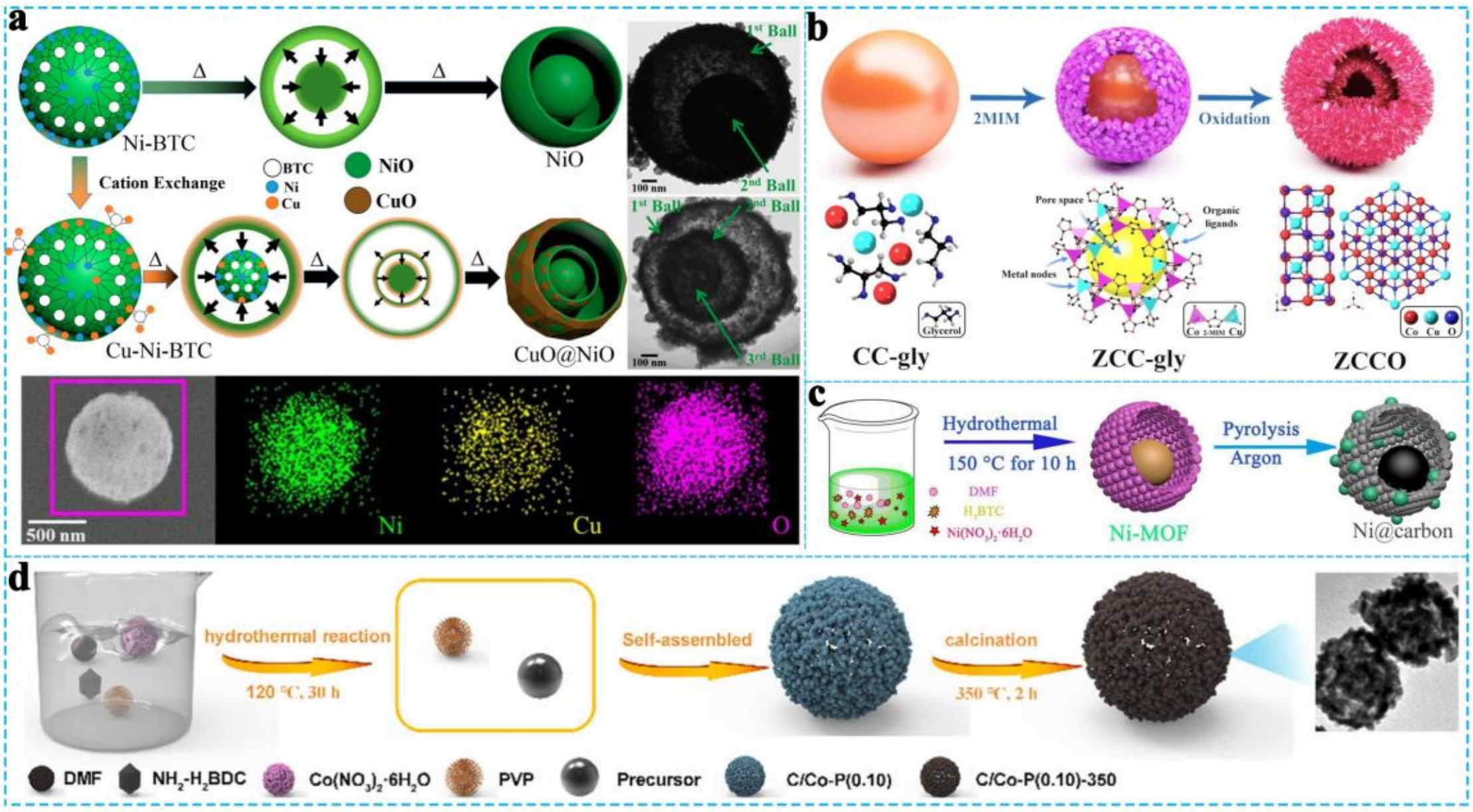

**Fig. 11.** The MOFs as self-templated for the synthesis of SMOs: (a) CuO@NiO hollow spheres by the microwave thermal treatment of Cu-Ni bimetallic organic frameworks. Reprinted with permission from Ref. [87]. (b) Double-shell $CuCo_2O_4$ hollow spheres by oxidizing CuCo-ZIF. Reprinted with permission from Ref. [88]. (c) Ni@carbon with 3D yolk shell-like structure by pyrolyzing Ni-MOFs. Reprinted with permission from Ref. [89]. (d) $Co_3O_4$ microsphere by calcining C/Co-MOFs. Reprinted with permission from Ref. [90].

Briefly, the self-template approach has several unique advantages as follows: (1) relatively simple, reproducible, and low-cost production procedures with controllable shell thickness and uniformity; (2) a better understanding of the chemical behavior of nanomaterials by involving interesting chemical transformation reactions; and (3) eliminating the need for heterogeneous coatings making it easier to scale up synthesis for mass production. However, it needs to face some disadvantages such as limited additives, high energy consumption for the removal of additives and considering the nature of the target structure. It has been regarded as an efficient strategy for the synthesis of SMOs with multi-shelled hollow structures by various self-template methods (**Table 3**). The hollow structures have a large surface volume ratio, high porosity, and internal void space, which bring many benefits for providing large accessible active sites for catalytic degradation reactions. In addition, thin shells and cavities can reduce the resistance of electrons and holes to the target pollutants and enhance the mass transfer efficiency of the photocatalytic system.

**Table 3** Synthesis of SMOs by different self-template approaches.

| Contents | Method | Additive | Additive removal | Morphology | Refs. |
| --- | --- | --- | --- | --- | --- |
| $MCo_2O_4$ (M=Ni, Zn, Mn) | Self-template | MCo-glycerate spheres (M=Ni, Zn, Mn) | Heating | 1. Multilayer hollow sphere<br>2. SSA: 61.2 $m^2$ $g^{-1}$ ($NiCo_2O_4$) | [79] |
| $\alpha$-$Fe_2O_3$ | Self-template | Iron nitrate-sucrose composite microsphere | Annealing | 1. Multi-shelled hollow spheres<br>2. SSA: 17.3 $m^2$ $g^{-1}$ | [91] |
| Mixed metal oxide | Self-template | Metal-phenolic coordination polymers | Calcining | 1. Multi-shelled hollow spheres<br>2. SSA: 55-171 $m^2$ $g^{-1}$ | [92] |
| ZnO, $Al_2O_3$, $Co_3O_4$, $Fe_2O_3$, CuO | Self-template | Metal-phenolic coordination polymers | Calcining | 1. Mesoporous or hollow spheres<br>2. SSA: 70, 59 $m^2$ $g^{-1}$ (Zn, Co) | [93] |
| $MoO_2$/carbon composite | Self-template | Mo-glycerate | Annealing ($N_2$) | 1. Multilayer hollow spheres<br>2. SSA: 60.9 $m^2$ $g^{-1}$ | [94] |
| $NiCo_2O_4$, $ZnCo_2O_4$ | Self-template | XCo-glycerate spheres (X=Ni, Zn) | Annealing (air) | 1. Multilayer hollow spheres<br>1. SSA: 61.2 $m^2$ $g^{-1}$ ($NiCo_2O_4$) | [79] |
| $CoMn_2O_4$ | Self-template | MnCo-glycerate spheres | Annealing (air) | Yolk-shell spheres | [79] |
| $NiCo_2V_2O_8$ | Self-template | NiCo-glycerate, $VO_3^-$ ions | Annealing (air) | Yolk-double Shell Spheres | [95] |
| Co–Fe alloy/ N-doped carbon | Self-template | PS@dual-MOF | Calcining ($N_2$) | Hollow spheres | [96] |
| $ZnS/Co_9S_8$@ N-HCS | Self-template | Bimetallic organic framework (BIMMOF) | Pyrolyzing ($N_2$) | 1. Hollow carbon nanospheres<br>2. SSA: 299 $m^2$ $g^{-1}$ | [97] |
| $ZnCo_2O_4$ | Self-template | Zn/Co-ZIF-PS | Calcining | 1. Porous hollow spheres<br>2. SSA: 27.0 $m^2$ $g^{-1}$ | [98] |
| NiO/Ni@C | Self-template | Ni-MOFs | Pyrolyzing (Ar) | 1. Chestnut shell-like hollow sphere<br>2. SSA: 74.78 $m^2$ $g^{-1}$ | [99] |

### *2.2. Template-free method*

Due to the complex preparation steps and high cost of the template method, its application in the preparation of metal oxide microspheres is limited. As an alternative method, template-free approaches to synthesize metal oxide microspheres have also been developed, especially for constructing more complex structures. The template-free just means not adding excess template material in the synthesis process. Currently, mature mechanisms such as the Kirkendall effect, Ostwald ripening, and Oriented attachment are three main approaches for the template-free synthesis of SMOs [100-102].

The Kirkendall effect, which was named after metallurgist Ernest Oliver Kirkendall, is an interesting phenomenon found in the metallurgical process. It can be summarized as the process of vacancy formation during the diffusion of two metals due to different diffusion rates. In the process of nanocrystal formation, this effect will lead to the rapid diffusion phase being diffused outward from the core and then forming the hollow structure with a composite core-shell structure [100]. This was firstly demonstrated by the group of Alivisatos, for regulating the synthesis of hollow $Co_xS_y$, CoSe, and Pt@CoO structures [103], and then further demonstrated by many other groups [104-106]. For the synthesis of SMOs via Kirkendall diffusion, the key is to regulate the core diffusion phase to spread out faster.

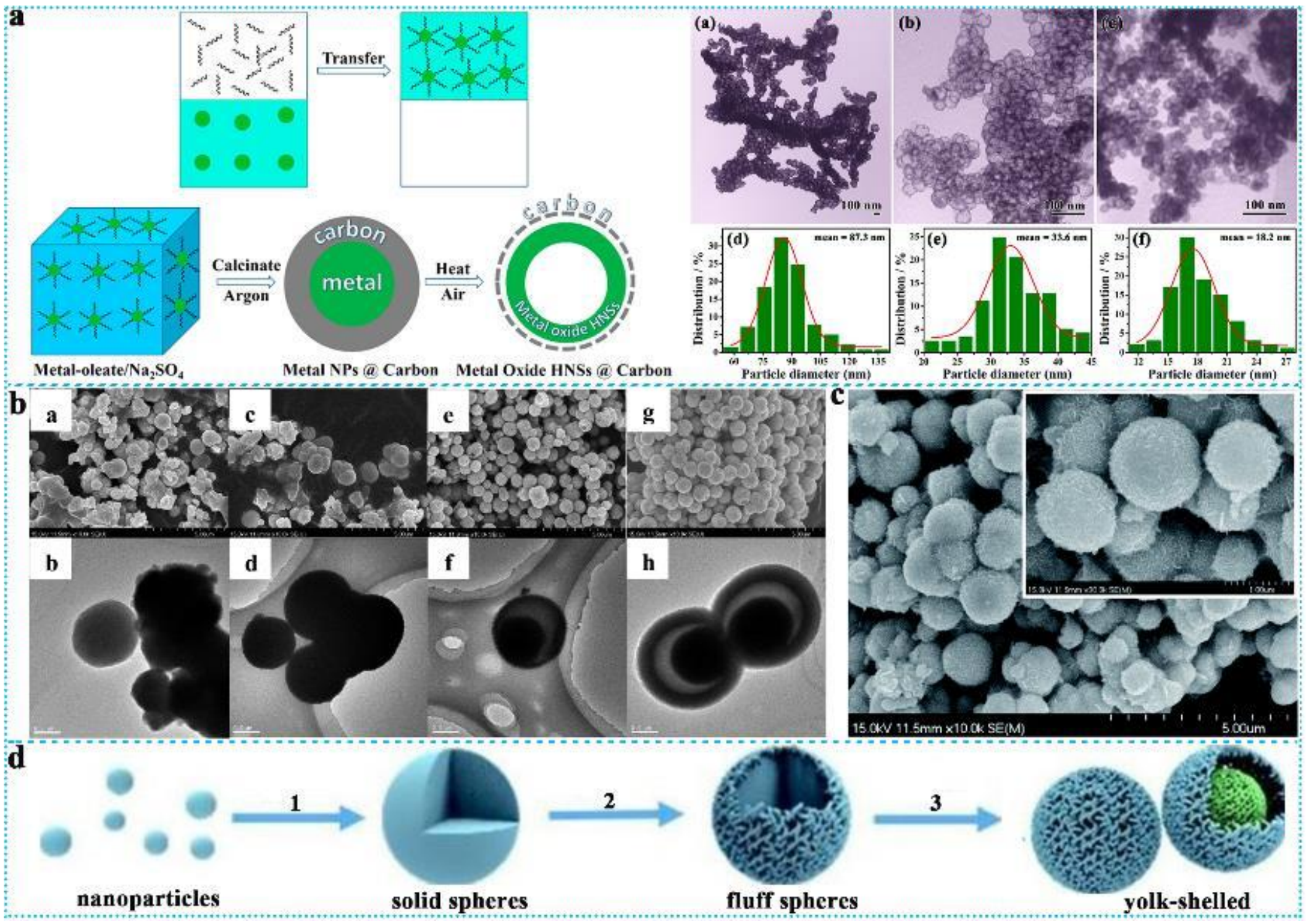

**Fig. 12.** The synthesis of SMOs by the Kirkendall effect: (a) CuO, NiO, and NiO/CuO

hollow nanospheres. Reprinted with permission from Ref. [107]. (b-d) yolk-shell structures $WVO_x$. Reprinted with permission from Ref. [108].

A variety of nickel oxide (NiO), copper oxide (CuO), and nickel-copper composite oxide (NiO/CuO) hollow spheres decorated with carbon shells were successfully prepared by the Kirkendall effect [107]. By further characterization (**Fig. 12a**), these three prepared catalysts showed great mesoporous nanosphere structure with average particle sizes of 87.3/33.6/18.2 nm, and the standard deviations were 12.7, 4.77, and 2.81 nm, respectively. In addition, small molecules could enter the internal space of the material due to the presence of mesoporous gaps, which is expected to increase the area of the active sites. Therefore, it showed excellent performance and catalytic stability in the hydrogenation reduction of 4-nitrophenol.

Shi et al. [108] established a simple synthesis method by combining the Ostwald ripening and the Kirkendall effect to synthesize a series of $WVO_x$ precursors with yolk-shell structures for further sulfurization. The morphology of $WVO_x$ formed at different growth times under hydrothermal at 200 °C was explored (**Fig. 12b-c**). With the increase of hydrothermal time (3, 6, 12, 24 h), the material gradually became a hollow spherical structure formed by vanadium surrounding tungsten. According to the mechanism of Ostwald ripening and the action mechanism of the Kirkendall effect, it was speculated that the material eventually formed a hollow structure due to the Ostwald ripening whereas the Kirkendall effect made the material eventually form vanadium enclosing tungsten (**Fig. 12d**).

Ostwald ripening, first described by William Ostwald in 1986, is a phenomenon observed in solid solutions or liquid sols. Crystal growth is shown as the gradual deposition of small particles onto larger particles, eventually forming a spherical shape, which meets the trend of minimum surface area (lowest energy) [109]. Although the phenomenon was discovered early, its application to the synthesis of hollow materials at the micro/nanoscale was not realized until 100 years later. Recently, Kang's groups [110] attempted to synthesize $FeO_x$ microspheres of $FeO_x$ and N-doped graphite ($FeO_x$-NGC/Y) with a yolk-shell structure through the Ostwald ripening mechanism (**Fig. 13a**). In the preparation process, dicyandiamide was decomposed with the release of heat, which promoted Ostwald ripening, resulting in the yolk-shell structure based on metal oxide materials. Then, N-doped graphite was derived from sucrose and dicyandiamide filled the gaps between the structures, and eventually formed a stable sphere. In the work of Zheng's groups[111], hollow $TiO_2$ spheres with high photocatalytic activity were prepared by a one-pot hydrothermal method without a template (**Fig. 13b**). The prepared $TiO_2$ catalyst completely degraded $1\times10^{-5}$ M RhB in 90 min, and the activity was considered to be derived from its hollow spherical structure with a larger surface area and more active sites exposed.

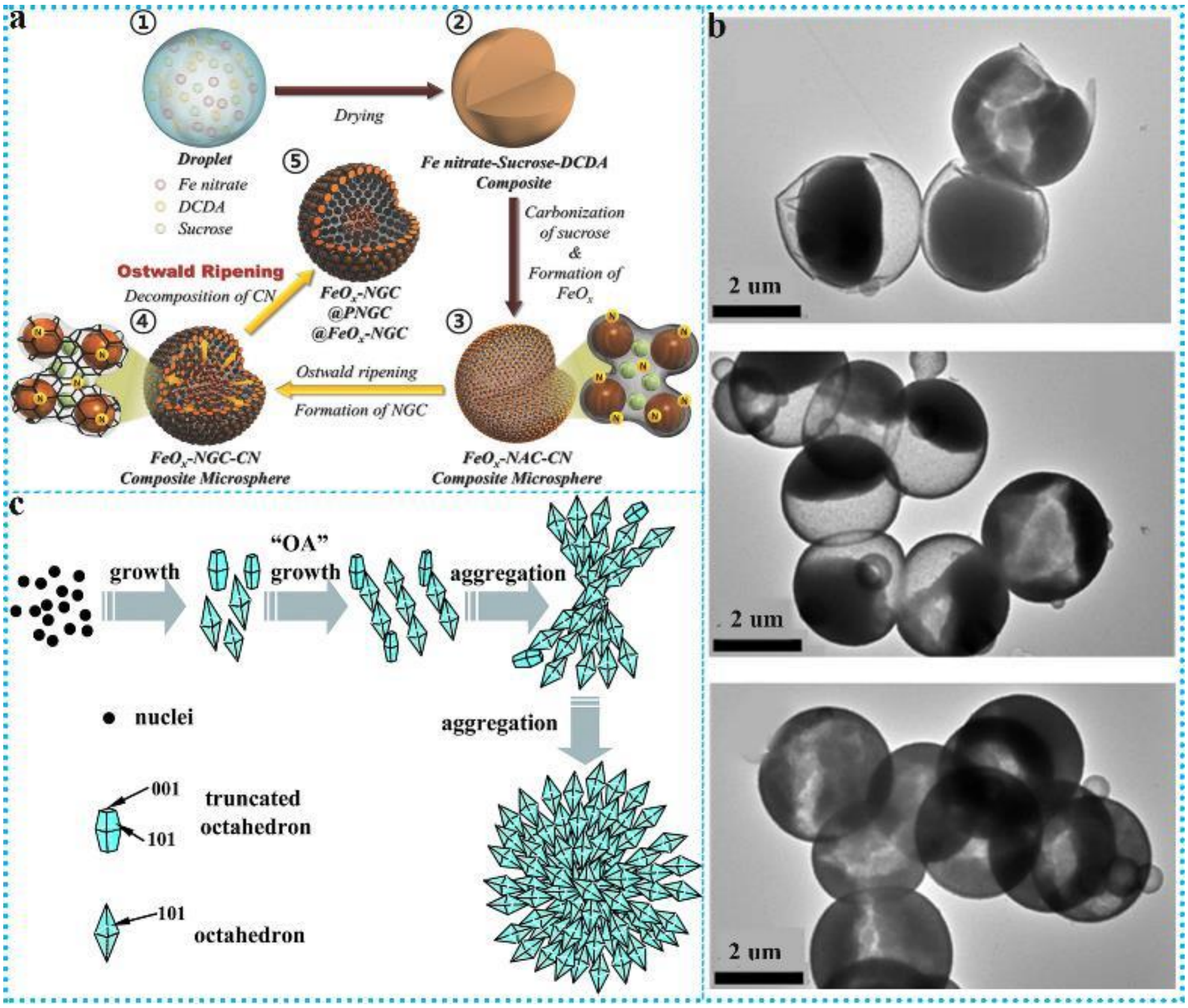

**Fig. 13.** The synthesis of SMOs by Ostwald ripening: (a) $FeO_x$-NGC/Y composite microspheres. Reprinted with permission from Ref. [110]. (b) hollow $TiO_2$ spheres. Reprinted with permission from Ref. [112]. The synthesis of SMOs by Oriented attachment: (c) the tentative mechanism for the formation of hierarchical $TiO_2$ microspheres. Reprinted with permission from Ref. [111].

Oriented attachment, mainly refers to the phenomenon that crystals aggregate along the same crystal orientation to form large crystals in the process of crystal growth, which can reduce the total surface area (reduce the surface energy). The morphology produced by this process mainly depends on the geometry and growth orientation of its original grains and the crystallization stability of the final product under this synthesis condition. Wei et al. [112] controlled the synthesis of hierarchical $TiO_2$ microspheres utilizing an oriented attachment growth model (**Fig. 13c)**. Besides, they studied the morphology of anatase $TiO_2$ varied with different reaction times. It was also observed that nanorods clumped together to form microspheres with porous structures when the sample was synthesized at 120 °C for 6 h. Similarly, Xiong et al. [113] designed a synthetic route of Ostwald ripening-assisted oriented attachment to the synthesis of $ZnMn_2O_4$ porous twin-microspheres. After adjusting the reaction parameters and time, the spherical precursor with a complete structure and $ZnMn_2O_4$ porous double microspheres with a diameter of 1 μm were successfully prepared.

**Table 4** shows other template-free approaches for the synthesis of SMOs. It is highly believed that template-free methods can promote the development of SMOs with the structure of core shells with the typical advantages such as (1) a wide range of applicable sizes and feasibility to form highly crystalline SMOs; (2) reasonably design SMOs with required activity and durability; and (3) encapsulating active nanomaterials in the preparation of yolk-shell/rattle-type structure containing active nanomaterials. Hence, such a multi-shell hollow structure of SMOs can not only protect the aggregation of internal nanospheres but also benefit the rapid diffusion of ROS. However, the template-free method has some disadvantages. For example, applying the method to small-scale particles remains challenging. In addition to metal oxides, the method can be extended to other novel functional materials, including metal-organic frameworks, covalent organic frameworks, and coordination polymers.

**Table 4** Synthesis of SMOs by different template-free approaches.

| Contents | Synthetic method | Precursors | Morphology | Refs. |
|---|---|---|---|---|
| $Co_3O_4$ | Kirkendall effect | $Co(NO_3)_2•6H_2O$ | 1. Hollow sphere<br>2. AD: 12.3 nm | [114] |
| β-$Bi_2O_3$ @$CeO_2$ | Kirkendall effect | $Bi_2O_3$ microspheres, $Ce(NO_3)_3•6H_2O$ | 1. Microsphere<br>2. AD: 1.87 µm | [115] |
| $Bi_2WO_6$ | Kirkendall effect | BiOBr solid microspheres | 1. Hollow microspheres<br>2. AD: 2-5 µm | [116] |
| $Bi_2S_3$ | Kirkendall effect | $Bi_2O_3$ microspheres, thioacetamide | 1. Hollow sphere<br>2. AD: 180 nm | [117] |
| $TiO_2$ | Ostwald ripening | $TiF_4$ | Nanospheres | [118] |
| $Cu_2O$ | Ostwald ripening | $Cu(Ac)_2•H_2O$, PVP, TEA | 1. Hollow sphere<br>2. AD: 400-700 nm | [119] |
| $NiCo_2O_4$/ NiO | Ostwald ripening | $Ni(NO_3)_2•6H_2O$, $Co(NO_3)_2•6H_2O$, $H_3BTC$ and PVP | 1. Hollow microspheres<br>2. AD: 1-3 µm | [120] |
| $NiCo_2O_4$ | Ostwald ripening | $Ni(NO_3)_2•6H_2O$, $Co(NO_3)_2•6H_2O$, Urea | 1. Hollow nanoparticles<br>2. AD: 60-100 nm | [121] |
| $CeO_2$ | Oriented attachment | $CeCl_3•7H_2O$, PVP | 1. Hollow spheres<br>2. AD: 0.3-4 µm | [122] |
| $Bi_2S_3$ | Oriented attachment | $Bi(NO_3)_3•5H_2O$, thiourea | Hollow spheres | [123] |
| $ZnMn_2O_4$ | Oriented attachment | $Zn(Ac)_2•2H_2O$, $Mn(Ac)_2•4H_2O$ | 1. Microspheres<br>2. AD: 1.0 µm | [93] |
| Black $TiO_2$ | Oriented attachment | Oxalic acid | 1. Mesoporous hollow spheres<br>2. AD: 35-115 nm | [124] |

SMOs with different morphologies and structures can be tuned by different template methods, which will be conducive to the performance of catalytic degradation. The pros and cons of SMO synthesis methods are thus summarized in **Fig. 14**. It is clear to see the benefits and drawbacks of these strategies.

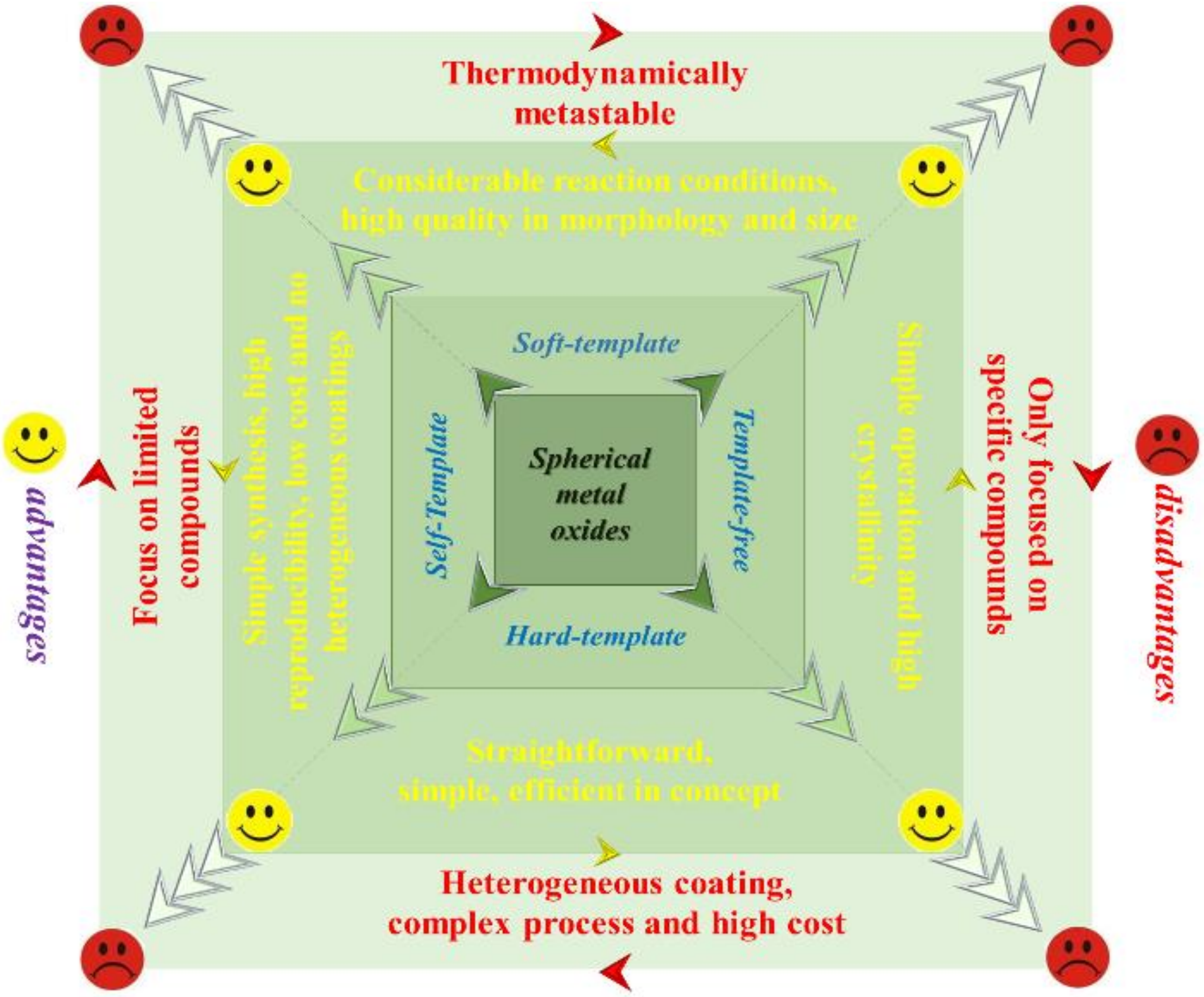

**Fig. 14.** Advantages and disadvantages of different preparation approaches for SMOs.

## 3. Performance and mechanism on the degradation of antibiotics over SMOs

### *3.1. General mechanism*

In the above review process, we have summarized the synthesis methods of SMOs regulating different morphologies. However, the link between morphology, catalytic activity, and the mechanism of SMOs is still unclear. As shown in **Fig. 15**, the photocatalytic process of SMOs generally includes four steps: (1) The generation of photogenerated $e^-/h^+$ pairs occurs when the light energy absorbed by SMOs semiconductor photocatalyst is equal to or greater than the band gap energy (**Eqn. 1**). (2) This pair of $h^+$ and $e^-$ move to the corresponding VB and CB, respectively. (3) On the photocatalyst surface of SMOs, these charge carriers in CB and VB can reduce $O_2$ and oxidize $H_2O$, respectively, thus forming the strong oxidizing ROS (•OH, $•O_2^-$ and $h^+$) (**Eqns. 2-8**). (4) Finally, the adsorbed antibiotics can be directly oxidized to small molecular intermediates and ideally mineralized into $CO_2$ and $H_2O$ by ROS (**Eqn. 9**).

$$\text{SMOs} + h\nu \rightarrow h^+_{VB} + e^-_{CB} \quad (1)$$

$$O_2 + e^-_{CB} \rightarrow •O_2^- \quad (2)$$

$$H_2O + h^+_{VB} \rightarrow H^+ + •OH \quad (3)$$

$$OH^- + h^+_{VB} \rightarrow •OH \quad (4)$$

$$•O_2^- + H_2O \rightarrow HO_2• + OH^- \quad (5)$$

$$2\ HO_2• \rightarrow O_2 + H_2O_2 \quad (6)$$

$$HO_2• + H_2O + e^-_{CB} \rightarrow H_2O_2 + OH^- \quad (7)$$

$$e^-_{CB} + H_2O_2 \rightarrow OH^- + •OH \quad (8)$$

$$\text{Antibiotics} + \text{ROS}\ (•OH + •O_2^- + h^+_{VB}) \rightarrow \text{intermediates} \rightarrow CO_2 + H_2O \quad (9)$$

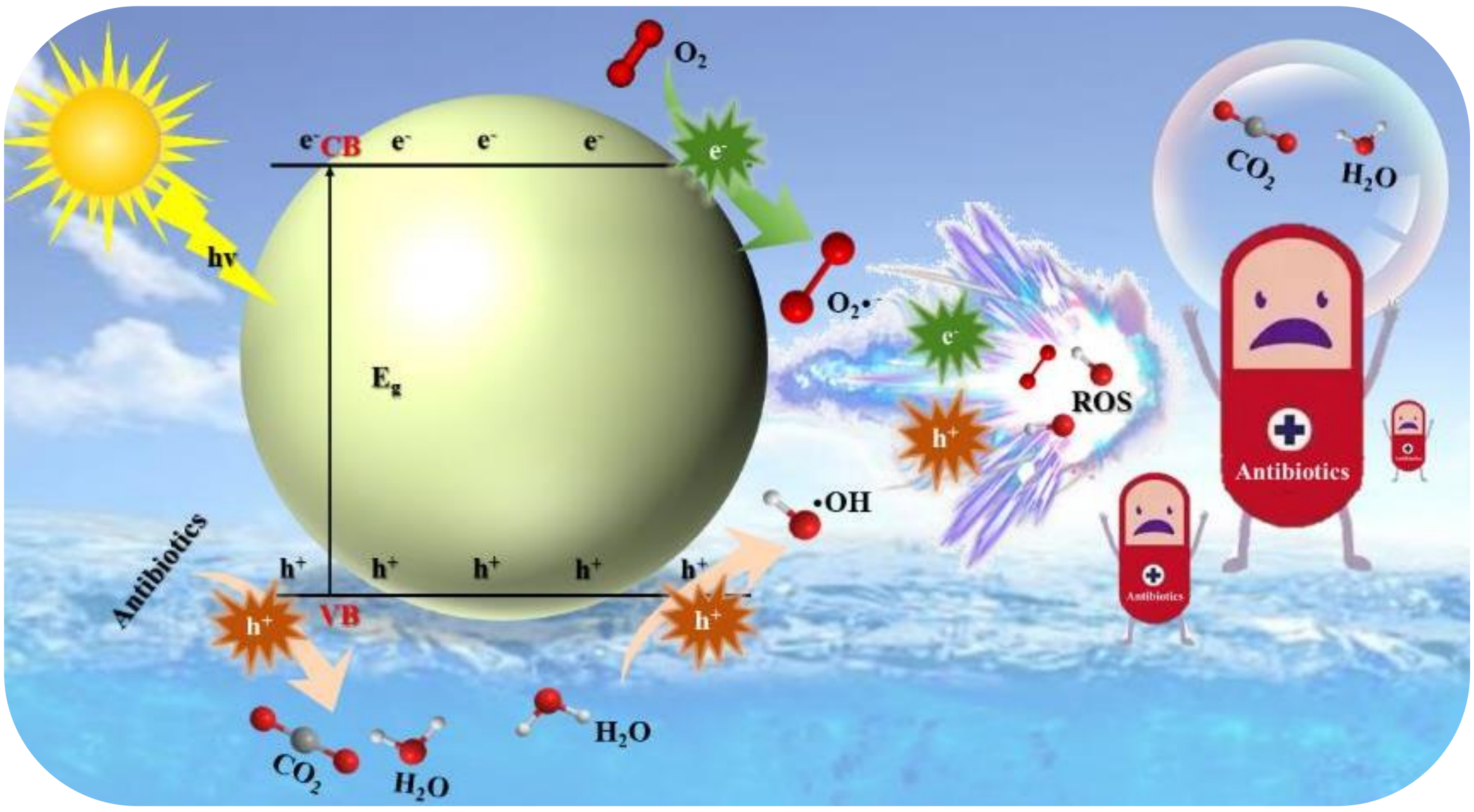

**Fig. 15.** The photocatalytic mechanism of a SMO semiconductor photocatalyst for antibiotics degradation.

To further improve the efficiency of photocatalysis with SMOs, inorganic semiconductor modification and applying a combined Fenton system are the most potential development directions. For example, our group proposed a photo-Fenton-like synergistic mechanism of the Ag/CNQDs@$Fe_3O_4$ hollow microspheres [125]. As shown in **Fig. 16**, it is found that SMOs with a hollow structure can make rational use of light energy. When the light shines into the interior of the ball, the energy can be reflected in the Ag/CNQDs@$Fe_3O_4$ constantly, which makes it easier to generate electrons by light. Besides, after adding persulfate into the photocatalytic system (see **Eqns. 10-11**), the system can not only alleviate the difficulty in photogenerated carrier separation during the photocatalytic process but also accelerate the activation process, thereby greatly promoting the generation of ROS and improving the removal efficiency of antibiotics.

$S_2O_8^{2-} + e_{CB}^{-} \rightarrow SO_4^{2-} + SO_4^{\bullet-}$ (10)

$HSO_5^{-} + e_{CB}^{-} \rightarrow OH^{-} + SO_4^{\bullet-} / SO_4^{2-} + {}^{\bullet}OH$ (11)

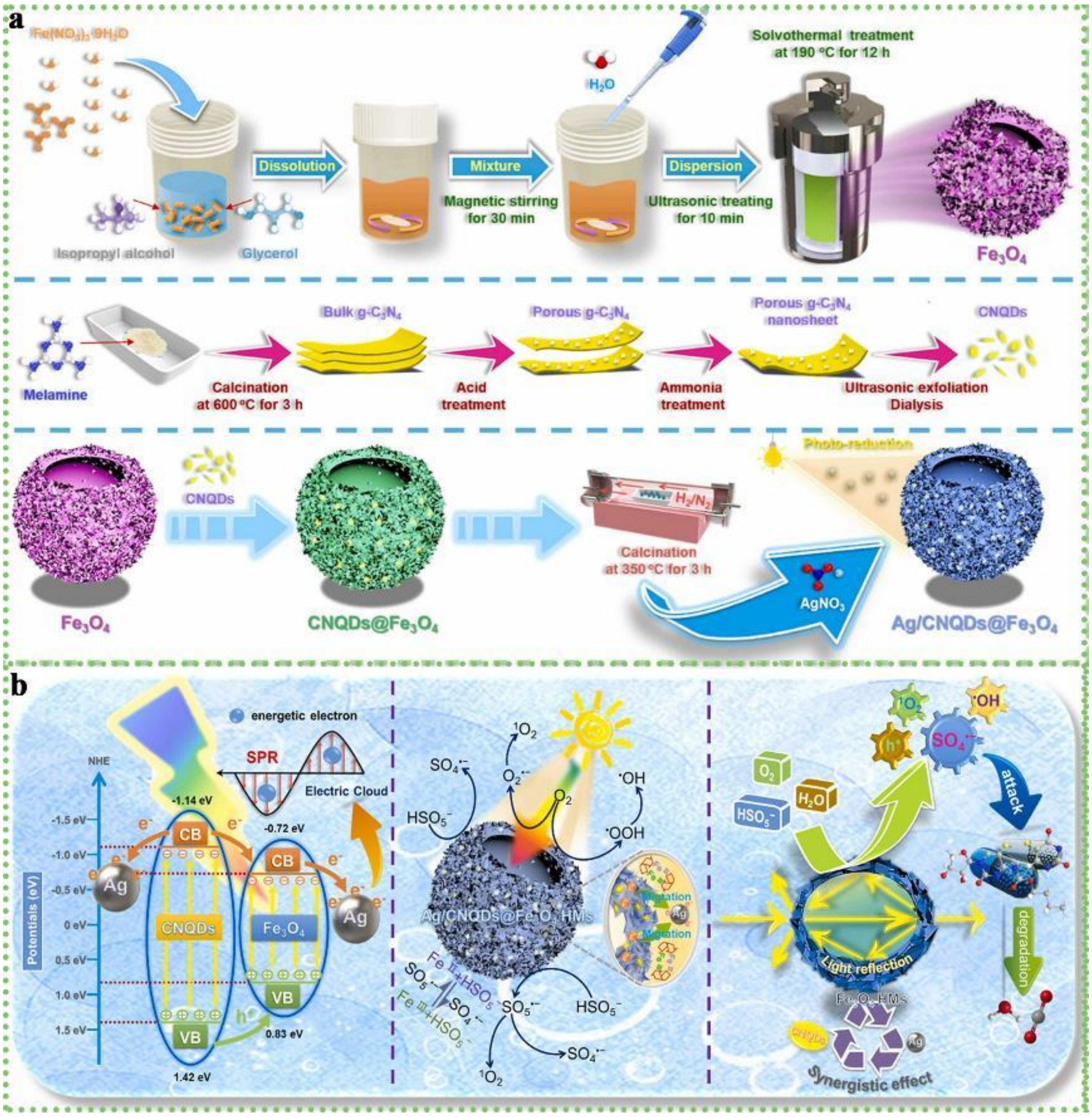

**Fig. 16.** Our group proposed a photo-Fenton-like synergistic mechanism of the Ag/CNQDs@$Fe_3O_4$ hollow microspheres. Reprinted with permission from Ref. [125].

Moreover, the morphology of other SMOs as a semiconductor is closely related to its photocatalytic mechanism: (a) SMOs with a porous structure can make light completely enter the active sites. At the same time, pollutants can be enriched in the pores, which can quickly react with the ROS generated inside. Hence, the system can continuously generate photo-generated electron $e^-/h^+$ pairs. (b) SMOs with a yolk-shell nanostructure with active nanomaterials encapsulated inside can not only protect the aggregation of internal nanospheres but also allow the effective diffusion of ROS, obtaining high quantization efficiency. (c) SMOs with a multi-shelled hollow structure can expose more catalytic active sites to enable good separation of $h^+$ and $e^-$ under light irradiation. Moreover, thin shells and cavities can reduce the resistance of $h^+$ and $e^-$, improving the production of ROS in the reaction system. For the antibiotic degradation of photocatalysts based on SMOs, researchers are committed to improving their morphology and performance, mainly by modifying and combining other semiconductors, to absorb longer wavelengths and avoid the recombination of charge carriers. Traditional semiconductor SMOs, such as $TiO_2$, ZnO, $SnO_2$, $Cu_2O$, $Fe_2O_3$, $CeO_2$, and $WO_3$, have received widespread attention due to their ability to degrade various organic pollutants [126]. **Table 5** summarizes the performance of common metal oxides for photocatalytic degradation of organic pollutants. However, these traditional SMOs have drawbacks such as a small absorption range of the solar spectrum, wide bandgap, and fast recombination speed of photogenerated electrons and holes, which reduces their photocatalytic activity [127, 128]. Therefore, we outline the catalytic performances and mechanisms of different types of SMOs for the degradation of antibiotics (tetracyclines, quinolones, sulfonamides, and other antibiotics) with the main photocatalyst involved being SMOs-based composites. Meanwhile, we have also discussed the mechanism of antibiotic degradation by SMOs alone.

**Table 5** Common metal oxides for photocatalytic degradation of organic pollutants.

| Metal oxides | Morphology | Target pollutants | Reaction conditions | Performance | Refs. |
|---|---|---|---|---|---|
| $TiO_2$ | Mesoporous spheres | 2-(4-methylphenoxy)ethanol (MPET) | [MPET] = 20 mg $L^{-1}$, [catalyst]= 0.4 g $L^{-1}$, UV light source (500 W high-pressure mercury lamp) | 100% (300 min) | [129] |
| $SnO_2$ | Triple-shelled hollow spheres | Methylene blue (MB) | [MB] = 10 mg $L^{-1}$, [catalyst]= 1.0 g $L^{-1}$, UV light source (250 W high-pressure mercury lamp) | 71.8 % (180 min) | [130] |
| $WO_3$ | Hierarchical hollow nest | TC | [TC] = 20 mg $L^{-1}$, [catalyst] = 1.0 g $L^{-1}$, visible light source (300 W Xe lamp) | 94.3 (60 min) | [131] |
| $Fe_2O_3$ | Hollow spheres | Salicylic acid (SLA) | [SLA] = 50 mg $L^{-1}$, [catalyst]= 0.4 g $L^{-1}$, UV light source | 74% (110 min) | [132] |
| ZnO | Porous spheres | Methyl orange (MO) | [MO] = 20 mg $L^{-1}$, [catalyst]= 0.2 g, UV light source | 96.3% (120 min) | [133] |
| $Cu_2O$ | Hierarchical hollow spheres | MO | [MO] = 20 mg $L^{-1}$, [catalyst] = 0.2 g $L^{-1}$, visible light source (250 W Xe lamp) | 98.5 % (120 min) | [134] |
| $Ir_2O_3$ | Porous spheres | Rhodamine B (RhB) | [RhB] = 5 mg $L^{-1}$, [catalysts] = 1.0 g $L^{-1}$, UV light source | 99 % (180 min) | [135] |
| $CeO_2$ | Hollow spheres | TC | [TC] = 20 mg $L^{-1}$, [catalyst] = 0.5 g $L^{-1}$, visible light source (300 W Xe lamp) | 100 % (60 min) | [136] |

### *3.2. Tetracycline antibiotics*

Tetracyclines (TCs) are the specific types of these drugs including tetracycline, chlortetracycline, oxytetracycline, and some semi-synthetic derivatives such as doxycycline, etc. [137]. TCs contain functional groups such as a hydroxyl group, an enol hydroxyl group, and a carbonyl group, and thus their properties are unstable under acid or base conditions. However, they can combine with metal-related ions to form chelates under neutral conditions [138].

Due to the wide application of TCs in the treatment of human diseases and animal breeding and medical production, they have a large number of residues in the water and soil environments, which have brought many adverse effects to the ecosystem, resulting in the emergence of multi-resistant bacteria and greatly affecting the life activities of aquatic animals. The main ways for TCs antibiotics to enter the environment are animal breeding, medical industry, and industrial discharge. After being used by humans or animals, these drugs cannot be completely absorbed in two final ways: one part is metabolized in the body and finally converted into low-active compounds, and the other party retains the original drug or forms active metabolites, and finally passes through the urine, feces, etc. to discharge into surface water or groundwater, and is finally injected into the water environment [139]. In addition, the soil is also a major destination for TCs antibiotics. Drugs enter the soil through medical systems or industrial wastewater discharge, etc., causing pollution to the soil environment. During the planting process, it is inevitable that residual drugs will enter the plant body from the soil, and ultimately affect human health and even life safety through food [140]. Therefore, if TCs are not disposed of properly, they will have a non-negligible impact on the environment.

Tang et al. [141] prepared a rambutan-shaped hollow $ZnFe_2O_4$ (ZFO) sphere photocatalyst composed of slender nanoparticles with different thicknesses (20-50 nm). The calcination temperature had an effect on the photocatalytic performance of ZFO.

The ZFO-400 exhibited the best photocatalytic activity based on activating persulfate (PDS), which could degrade 100% of TC in 60 min by $SO_4^{\bullet-}$, •OH, $\bullet O_2^-$ and $h^+$. Although the $ZnFe_2O_4$ alone has good photocatalytic performance, additional PDS oxidants are required in the system to improve degradation efficiency, thereby increasing the complexity and cost of the system. To improve the photocatalytic performance of SMOs alone, Huang et al. [142] reported a series of 3D layered bismuth oxyiodides prepared using a self-sacrificed template (BiOI microspheres) through an in-situ phase transformation and phase structure building process. The obtained $Bi_4O_5I_2$-$Bi_5O_7I$ catalyst had a good photocatalytic performance for removing many different antibiotics and pollutants such as TC, bisphenol A (BPA), and rhodamine B (RhB). Especially, $Bi_4O_5I_2$-$Bi_5O_7I$ could degrade 78% TC within 1 h under visible light irradiation ($\lambda > 420$ nm) due to its clear microstructure and energy band structure, which was almost 3 times that of the BiOI. No obvious performance decrease for the TC degradation were observed after 4 cycles, suggesting that the $Bi_4O_5I_2$-$Bi_5O_7I$ photocatalyst had strong stability and good recyclability for TC degradation.

Importantly, Yu et al. [136] found that $CeO_2$ hollow spheres (CHS) had a size effect on photocatalytic performance for TC degradation. As the annealing temperature increased from 600 to 800 °C, the size and average pore size of CHS significantly increased. On the contrary, the specific surface area decreased. When the calcination temperature was 600 °C, CHS-600 had a smaller size and average pore size ( A spherical shell thickness of 30 nm and an inner diameter of 650 nm) and a larger specific surface area (58 $m^2/g$) compared to CHS-700 and CHS-800. As expected, the as-obtained CHS-600 exhibited higher photocatalytic activity to degrade TC, which was attributed to that the size effect of hollow spheres could not only effectively accelerate the multiple reflections of light in the inner cavity and improve the separation of charge carriers, but also provide a larger specific surface area and abundant surface active sites.

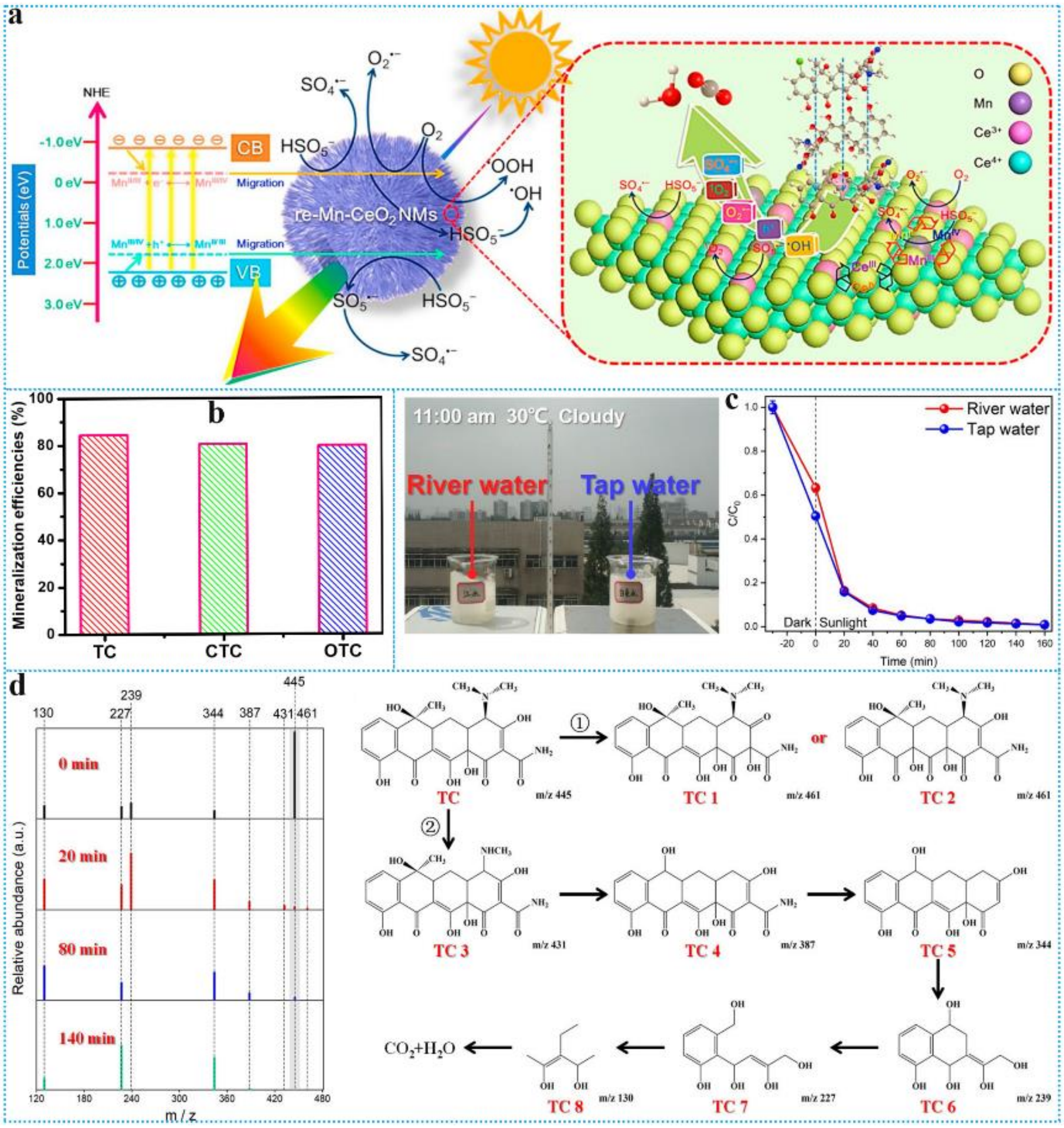

**Fig. 17.** (a-b) Our group proposed degradation mechanisms of TC in the re-Mn-$CeO_2$ NMs microflowers/PMS/vis system. Reprinted with permission from Ref. [75]. The removal rate (c), ESI-MS signals of intermediates and degradation pathways of TC (d) in the system of HTB hollow spheres/vis. Reprinted with permission from Ref. [143].

Except for the hollow particles made up of a single material, hollow structures formed by various components have also attracted much attention from academia. For example, our group [75] reported re-Mn-$CeO_2$ NM as an excellent Fenton-like photocatalyst for removing various TCs synergistically including tetracycline (TC), chlortetracycline (CTC), and oxytetracycline (OTC) in a coupling system consisting of sulfate radical like Fenton process and visible light driven photocatalysis (**Fig. 17a**). Especially, when the reaction system was the optimal (re-7Mn-$CeO_2$ NMs/PMS/Vis), the degradation efficiencies of TC, CTC and OTC reached 98.6%, 97.4%, and 88.1% with the mineralization efficiencies of 84.3%, 80.6%, and 79.8% respectively (**Fig. 17b)**. Shi et al. [143] successfully prepared the hollow sphere $TiO_2/Bi_2O_3$ photocatalyst (HTB), which exhibited great practical potential in wastewater treatment. **Fig. 17c** shows that TC in tap water and river water can be completely removed when the HTB system is used in outdoor light (cloudy), suggesting that the HTB had a satisfactory photocatalytic performance for treating actual wastewater. Besides, the main two TC degradation pathways in the system of HTB/vis are illustrated in **Fig. 17d**. The •OH attacked different positions of TC molecules to obtain the structure of TC1 or TC2 via a hydroxylation process (path I). Path II was demethylation and double hydroxylation processes. TC3 was converted to TC4 through the attack path of $•O_2^-$ and $h^+$. Then, remove the amide group and transfer TC4 to TC5 through deamidation. TC6 and TC7 were formed through the continuous attack TC5 by $•O_2^-$ and $h^+$ via carbonylation and addition reaction. TC3-7 were gradually transformed into TC8 through ring-opening reaction. Finally, the intermediates of TC were mineralized into $H_2O$ and $CO_2$. However, the drawback is that the author did not evaluate the toxicity of these TC intermediates.

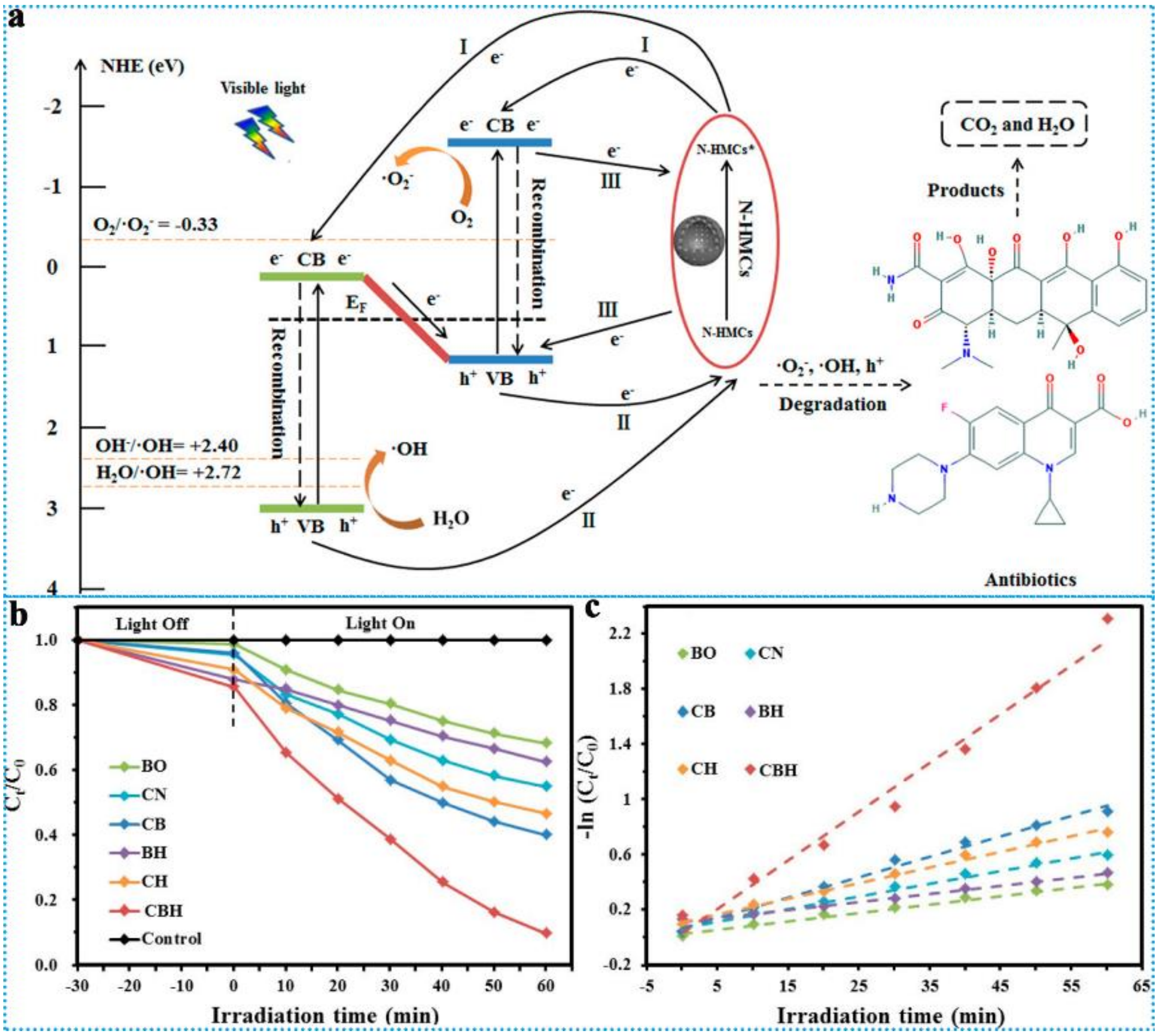

**Fig. 18.** Photocatalytic mechanism (a) and photocatalytic activity (b) for TCH degradation under CBH/visible-light (λ > 420 nm) system. (c) Pseudo-first-order kinetic models for the photodegradation of TCH. Reprinted with permission from Ref. [144].

Zeng et al. [144] successfully prepared a novel visible-light photocatalyst, which is composed of nitrogen-doped hollow mesoporous carbon spheres (N-HMCs) and g-$C_3N_4$/$Bi_2O_3$. The dual semiconductor photocatalyst g-$C_3N_4$/$Bi_2O_3$@N-HMCs (CBH) was prepared by a simple thermal method. **Fig. 18a** shows that N-HMCs play three important roles in enhancing photocatalytic efficiency. First, N-HMCs could be used as photosensitizers to increase visible light absorption. Secondly, N-HMCs were excellent conductive materials, which could rapidly transfer photoexcited electrons and hinder the recombination of electron-hole pairs. Third, N-HMCs could provide more active sites in the photocatalytic degradation process. Therefore, it can be seen from **Fig. 18b-c** that the photocatalytic efficiency of the CBH complex for the degradation of tetracycline hydrochloride (TCH) is the best. At the same time, the photodegradation efficiency of single-phase $C_3N_4$ (CN), single-phase $Bi_2O_3$ (BO), binary composites g-$C_3N_4$/N-HMCs (CH), $Bi_2O_3$/N-HMCs (BH), and g-$C_3N_4$/$Bi_2O_3$ (CB) are superior to the corresponding single-phase samples g-$C_3N_4$ and $Bi_2O_3$. The application of other SMOs in the degradation of TCs is shown in **Table 6.**

**Table 6** Application of SMOs in the degradation of TCs.

| Catalysts | Morphology | Target pollutant | Reaction conditions | Performance | Mechanism | Refs. |
|---|---|---|---|---|---|---|
| $TiO_2$ | Mesoporous dark brown nanospheres | OTC | • 5 mg catalyst, 0.5 mg mL$^{-1}$ OTC solution<br>• Natural sunlight (Intensity ~ 925 w/cm$^2$) | 100% (80 min) | •$O_2^-$ and •OH | [145] |
| $ZnFe_2O_4$ | Hollow spheres | TC | • 5 mg catalyst, 10 mg L$^{-1}$ TC solution, 2 g L$^{-1}$ PDS<br>• Visible light (λ > 420 nm) | 100% (120 min) | $SO_4^{•-}$, •OH, •$O_2^-$and h$^+$ | [141] |
| $MnCeO_x$ | Hollow spheres | TC | • 20 mg catalyst, 20 mg L$^{-1}$ TC solution,<br>• Visible light (λ > 420 nm) | 75% (90 min) | •OH and •$O_2^-$ | [146] |
| $SiO_2$-$Fe_2O_3$@$TiO_2$ | Hollow spheres | TC | • 10 mg catalyst, 50 mL TC (10 mg L$^{-1}$)<br>• Visible light (λ > 420 nm) | 80% (80 min) | / | [61] |
| | | | • Simulated solar light (with an AM 1.5 G filter) | 100% (140 min) | h$^+$ | |
| | | | Natural sunlight: Sunny | 100% (80 min) | / | |
| | | | Natural sunlight: Cloud | 100% (140 min) | / | |
| $Ag/Ag_2CO_3$ $/BiVO_4$ | Flower-like sphere | TC | • 20 mg catalyst, 50 mL TC (20 mg $L^{-1}$) solution<br>• Visible-light illumination (500 W Xe lamp) | 94.9% (150 min) | •OH and $h^+$ | [147] |
| $Bi_2MoO_6$ | Hollow mesoporous nanostructures | TC | • 50 mg catalyst, 50 mL TC (10 mg $L^{-1}$) solution<br>• UV light irradiation (300W Hg vapor lamp) | 92% (90 min) | $•O_2^-$ and $h^+$ | [148] |
| $Bi_2MoO_6$ | Mesoporous spheres | TC | • 50 mg catalyst, 50 mL TC (30 mg $L^{-1}$) solution<br>• UV light irradiation (300W Hg vapor lamp) | 73% (150 min) | $•O_2^-$ and $h^+$ | [149] |
| $SrTiO_3$ (La, Cr)-6 | Spheres | TC | • 50 mg catalyst, 100 mL TC (20 mg $L^{-1}$) solution<br>• Visible-light illumination with 300W Xe lamp with a UV cutoff filter ($\lambda \geq 420$ nm) | 83% (90 min) | $•O_2^-$, •OH and $h^+$ | [150] |
| Mn-doped $CeO_2$ | Microflowers | TC<br>CTC<br>OTC | • 5 mg catalyst, 50 mL TC (30 μM) solution<br>• Visible light (500 W Xe lamp, $\lambda > 420$ nm, 100 mW $cm^{-2}$) | 100 % (60 min) | $SO_4^{•-}$, •OH,$^1O_2$ $•O_2^-$and $h^+$ | [75] |
| CT-g-ZnTAPc | Hollow spheres | TC• HCl | • 15 mg catalyst, 30 mL TC•HCl (20 mg $L^{-1}$) solution<br>• Visible light (300 W Xe lamp, $\lambda > 420$ nm) | 100 % (100 min) | $e^-$, $•O_2^-$ and $h^+$ | [151] |
| $g-C_3N_4/$ $Bi_2O_3$ @N-HMCs | Hollow mesoporous | TCH | • 100 mg catalyst, 100 mL TCH (10 mg $L^{-1}$) solution<br>• Visible-light irradiation (300 W Xe lamp, $\lambda > 420$ nm) | 90.6 % (60 min) | •OH, $•O_2^-$ and $h^+$ | [152] |
| $\gamma-Fe_2O_3/$ $b-TiO_2$ | Hollow sphere heterojunctions | TC | • 30 mg catalyst, 100 mL TC (10 mg $L^{-1}$) solution<br>• Simulated solar light (300 W Xe lamp with an AM 1.5 G filter, 100 mW $cm^{-2}$) | 99.3 % (50 min) | •OH, $•O_2^-$ and $H_2O_2$ | [153] |
| $WO_3$/ $g$-$C_3N_4$ | Hollow microspheres composite | TC •HCl | • 50 mg catalyst, 100 mL TC-HCl (25 mg $L^{-1}$) solution • Visible-light irradiation (300 W Xe lamp, λ > 420 nm) | 82 % (120 min) | •OH and $h^+$ | [154] |
| AgBr/ $Bi_2WO_6$ | Hierarchical flower-like | TC | • 50 mg catalyst, 50 mL TC (20 mol $L^{-1}$) solution • Visible-light irradiation (300 W Xe lamp, λ > 420 nm) | 87.5% (60 min) | •OH, •$O_2^-$ and $h^+$ | [155] |
| POPD-$CoFe_2O_4$ | Hollow sphere heterojunctions | TC | • 50 mg catalyst, 100 mL TC (20 mol $L^{-1}$) solution • visible light irradiation (PLS-SXE300, power 300 W, 1.8×105 lux) | 100 % (60 min) | $e^-$ and $h^+$ | [156] |
| $BiVO_4$/ $FeVO_4$@ rGO | Sphere heterojunctions | TC | • 30 mg catalyst, 50 mL TC (30 mg $L^{-1}$) • Visible light irradiation (1 KW Xe lamp, λ > 420 nm) | 91.5 % (100 min) | •$O_2^-$ and $h^+$ | [157] |

### *3.3. Quinolone antibiotics*

Fluoroquinolones (FQs) antibiotics belong to the third generation of quinolone antibacterial drugs, also known as pyridone acids, which are chemically-synthesized antibacterial drugs and have been mass-produced and applied since the 1970s [158]. FQs have fluorine atoms, nitrogen atoms, carboxyl groups, ketone groups, and a piperazine ring, so they are weakly acidic and basic. The typical drugs of this class of FQs are ciprofloxacin (CIP), ofloxacin (OFL), levofloxacin (LEV), and enrofloxacin (ENR). They have low solubility in water and are relatively stable to external stimuli such as light and heat [159], which makes a large number of residues in the water environment. However, the existing sewage treatment technology cannot completely remove them, and the drugs discharged into the environment will affect environmental organisms due to enrichment, and then endanger human health. Therefore, their environmental pollution problems have attracted widespread attention [160].

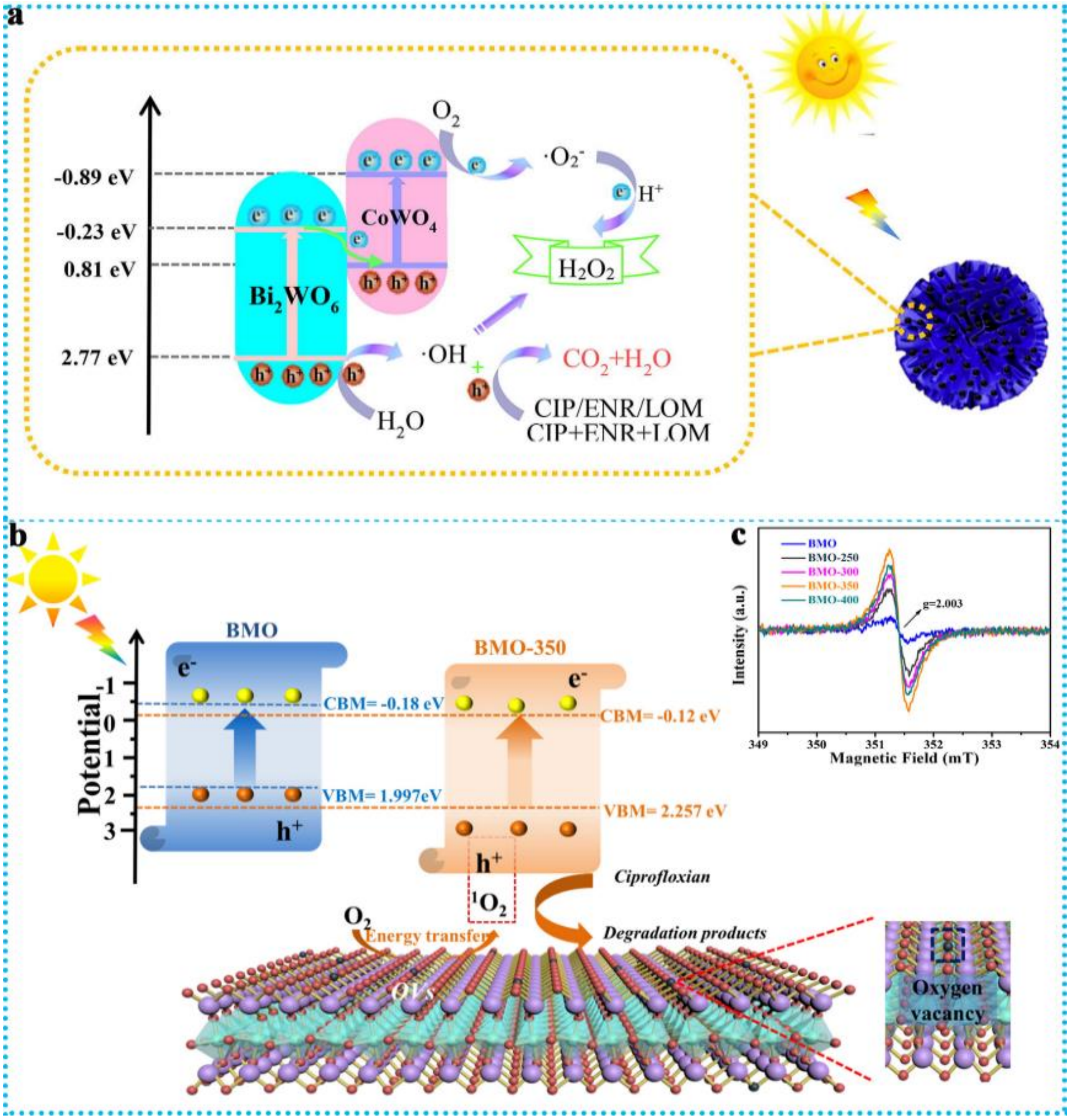

**Fig. 19.** Photocatalytic degradation of FQs: (a) 3D hierarchical micro-flowers $CoWO_4@Bi_2WO_6$ for CIP photocatalytic degradation. Reprinted with permission from Ref. [161]. (b) and (c) $Bi_2MoO_6$ double-layer spheres for visible light photocatalytic degradation of CIP. Reprinted with permission from Ref. [162].

Wang et al. [161] prepared $CoWO_4@Bi_2WO_6$ Z-scheme heterojunction for photocatalytic degradation for antibiotics degradation (**Fig. 19a**). By a facile solvothermal method, 3D hierarchical microflowers were constructed, largely increasing the specific surface area, which benefited the loading of $Bi_2WO_6$ nanoparticles on $CoWO_4$ for heterostructure construction, thus providing more active sites for pollutant degradation. Under visible light irradiation, removal efficiencies for ENR, LOM, and CIP achieved 84.6%, 83.6%, and 74.2% after 180 min, respectively, and photocatalytic degradation of the FQs mixtures (ENR-LOM-CIP) was also tested, and the total removal efficiency also reached 81.1% after 180 min.

Dong et al. [162] synthesized $O_V$-modified $Bi_2MoO_6$ double-layer microspheres for photocatalytic degradation of CIP (**Fig. 19b-c**). 90.6% of CIP was removed within 80 min under visible light. Double layer structure offered more active sites and allowed multiple light reflections inside microspheres, enhancing light harvesting. The introduction of oxygen vacancies increased valance band potential from 1.997 V vs. NHE to 2.257 V vs. NHE. Photogenerated holes with stronger oxidation ability could adsorb CIP molecules directly and efficiently degrade CIP. In addition, Zhang et al. [163] regulated the size effect of photocatalysts by self-assembly of different types of spheres. Sn-BiOI/ZnO (SBZs) multi-shelled microsphere was composed of ZnO microspheres with a diameter of about 600 nm and Sn-BiOI nanoflowers with a diameter of 1 μm. Although SBZs maintained the size of ZnO, there was a 200 nm hollow sphere inside its hollow sphere. It was found that the photocatalytic degradation rates of CIP by ZnO and BiOI were 72.7% and 74.6% under simulation sunlight irradiation within 100 min, respectively. For SBZs, CIP could be removed over 90%, far higher than that of pure ZnO and BiOI. The results indicated the multi-shelled hollow sphere structure of SBZs enhanced the departure of photogenerated electrons and holes, improving the utilization of light magnificently.

SMOs also exhibit great performance in Fenton or Fenton-like degradation for FQs. Zhang et al. [164] used MOF@$SiO_2$ as a precursor to design a yolk-shell $Co_3O_4$-C@$CoSiO_x$ Fenton-like catalyst with PMS as an oxidant for CIP degradation. 98.2% of CIP was efficiently removed within 17 min. This large activity was due to the high surface area of the yolk-shell structure, abundant highly active Co-$OH^+$, rich oxygen vacancies, and N-doped carbon for PMS activation. Results showed that the reaction includes a radical process and a non-radical process. Besides, density functional theory (DFT) calculations were carried out to predict the reaction sites of CIP molecules. The positions of CIP atoms, the highest occupied molecular orbital (HOMO), and the lowest unoccupied molecular orbital (LUMO) are shown in **Fig. 20a**, which indicated that N28, N38, O7, C4, C2, C10, C11, and N21 atoms as well as the ring were the main electron-rich sites in CIP molecules. This was confirmed by the degradation intermediates of CIP, which were mainly produced after these bonds were cracked. Finally, **Fig. 20b** shows the possible degradation pathways of CIP over $CoWO_4$@$Bi_2WO_6$ by using liquid chromatography-mass spectrometry (LC-MS) to analyze the intermediates of CIP degradation [161]. However, the relationship between ROS and the degradation pathway of CIP was not discussed in detail. In recent years, our team has developed a series of SMOs for the degradation of FQs. Wang et al. [78] studied mesoporous $MnO_x$ microspheres (**Fig. 20c**) under light conditions to activate PMS to form a photo-Fenton-like synergistic system. The influence of surface properties such as morphology, structure, and composition of the obtained materials on their catalytic performance was investigated through systematic characterization analysis. Based on LC-MS analysis, three main reaction pathways for CIP degradation including the cleavage of quinolone and piperazine rings, defluorination, and hydroxylation were proposed in the $MnO_x$/PMS/UV system. Pathway I: the piperazine ring of CIP was firstly broken by ROS. Pathway II: The F group of CIP was displaced by •OH and The C=C of the quinolone group was easily attacked by ROS. Finally, the ring-opening product was obtained through further benzene epoxidation and loss of the piperazine moiety. Pathway III: In the electron-rich region of CIP, such as the C=C bond of the benzene ring, hydroxylation was carried out by • OH. A synergistic photo-Fenton-like enhancement mechanism was found for the proposed system. **Table 7** displays the application of SMOs for the degradation of FQs.

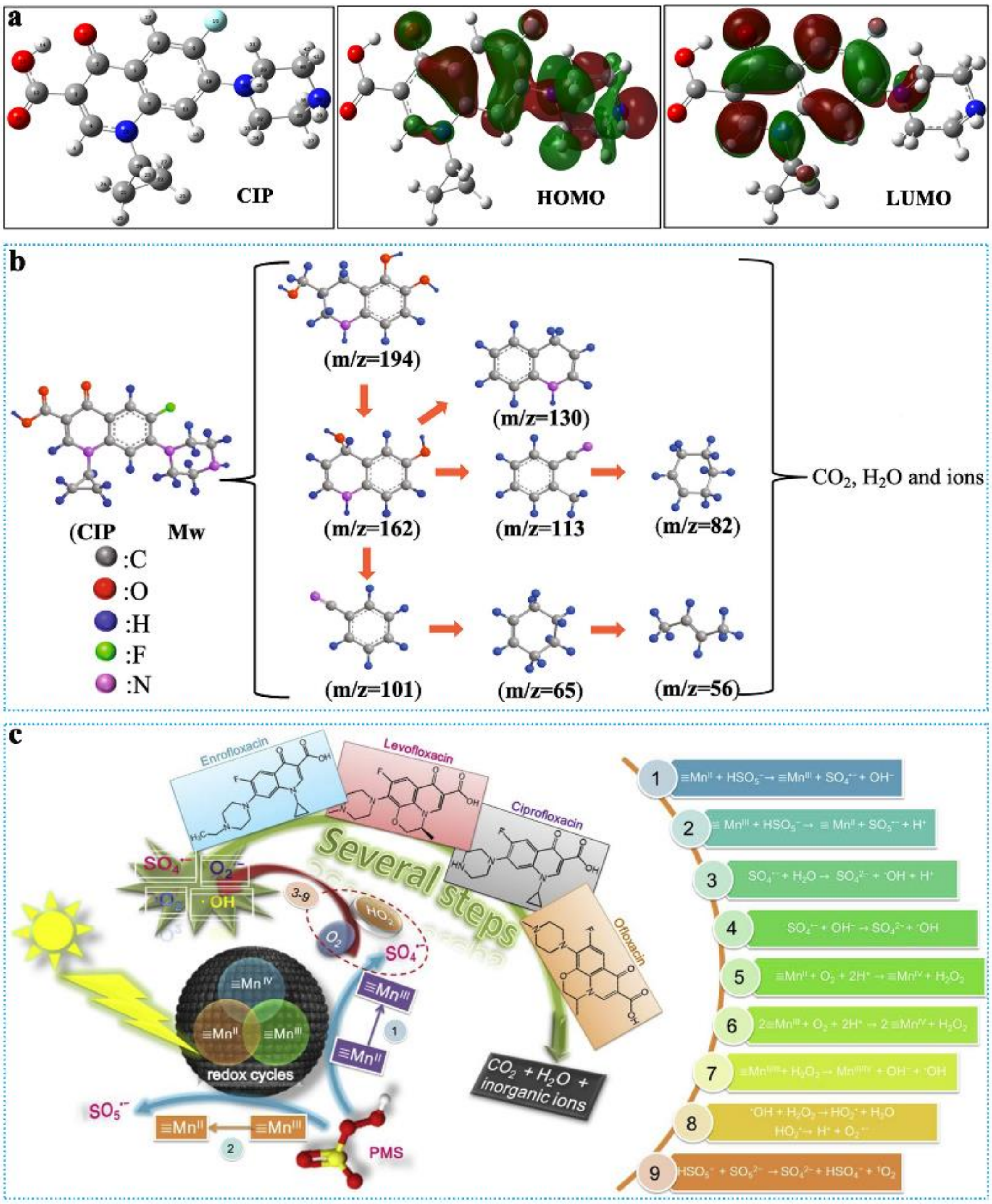

**Fig. 20.** (a) CIP atom site for the frontier electron densities calculation (HOMO and LUMO). Reprinted with permission from Ref. [165]. (b) The proposed main intermediates of the photodegradation of CIP via $CoWO_4@Bi_2WO_6$ microflowers under visible-light (λ > 420 nm) irradiation. Reprinted with permission from Ref. [161]. (c) Our group prepared mesoporous $MnO_x$ microspheres for photo-Fenton-like degradation of FQs. Reprinted with permission from Ref. [78].

**Table 7** Application of SMOs in the degradation of FQs.

| Catalysts | Morphology | Target pollutant | Reaction conditions | Performance | Mechanism | Refs. |
|---|---|---|---|---|---|---|
| $CoWO_4$ @$Bi_2WO_6$ | 3D hierarchical micro-flowers | LOM<br>CIP<br>ENR | • 40 mg catalyst, 50 mL pollutant (10 mg $L^{-1}$)<br>• Visible light (>420 nm, 300 W Xeon lamp) | 83.6% (180 min)<br>74.2% (180 min)<br>84.6% (180 min) | •OH, $h^+$ and •$O_2^-$ | [161] |
| $Bi_2MoO_6$ | Double-layer microspheres | CIP | • 50 mg catalyst, 80 mL CIP (10 mg $L^{-1}$)<br>• 18 W Visible LED light (35 mW $cm^{-2}$) | 90.6% (80 min) | $h^+$ and $^1O_2$ | [162] |
| Sn-BiOI /ZnO | Microspheres | CIP | • 20 mg catalyst, 200 mL CIP (10 ppm)<br>• Visible light (380-780 nm, 100 mW $cm^{-2}$) | >95% (100 min) | •$O_2^-$ and $h^+$ | [163] |
| $CuFe_2O_4$ | hollow sphere | CIP | • 0.5 g $L^{-1}$ catalyst, 40 mM $H_2O_2$, 50 mL CIP (10 mg $L^{-1}$) | 100% (30 min) | •OH | [165] |
| $Co_3O_4$-C@$CoSiO_x$ | Yolk-shell | CIP | • 50 mg catalyst, 20 mg $L^{-1}$ PMS, 100 mL CIP (30 mg $L^{-1}$) | 98.2% (17 min) | •$O_2^-$, $^1O_2$, $SO_4^{\bullet-}$ and •OH | [164] |
| $SiO_2$-$Fe_2O_3$@$TiO_2$ | Hollow spheres | ENR | 10 mg catalyst, 50 mL ENR (5 mg $L^{-1}$)<br>• Visible light ($\lambda > 420$ nm)<br>• Simulated solar light (with an AM 1.5 G filter)<br>• Natural sunlight<br>• Sunny cloud | 70.5% (160min)<br>100% (80 min)<br>100% (80 min)<br>98% (140 min) | $h^+$ with strong oxidation power | [61] |
| $WO_3$/g-$C_3N_4$ | Hierarchical hollow microspheres | CFS | 50 mg catalyst, 100 mL CFS (25 mg $L^{-1}$)<br>Visible-light irradiation (300 W Xe lamp, $\lambda >$ 420 nm) | 70 % (120 min) | •OH and $h^+$ | [154] |

**Table 7 (*continued*)**

| Catalysts | Morphology | Target pollutant | Reaction conditions | | Performance | Mechanism | Refs. |
|---|---|---|---|---|---|---|---|
| $MnO_x$ | Mesoporous microsphere | OFL | 0.1 g $L^{-1}$ catalyst, 30 μ M OFL | • UV (CEL-LAM500, 50 mW $cm^{-2}$, λ < 365 nm) | 99.5 % (10 min) | $•O_2^-$, $^1O_2$, $SO_4^{•-}$ and •OH | [78] |
| | | | | • Simulated solar irradiation (CEL-LAX500, wavelength range: 200-1000 nm, 100 mW $cm^{-2}$) | 74.5 % (10 min) | | |
| | | CIP | 0.1 g $L^{-1}$ catalyst, 15 μ M CIP | • UV (CEL-LAM500, 50mW $cm^{-2}$, λ < 365 nm) | 97.8 % (10 min) | | |
| | | | | • Simulated solar irradiation (CEL-LAX500, wavelength range: 200-1000 nm, 100 mW $cm^{-2}$) | 79.4 % (10 min) | | |
| | | ENR | 0.1 g $L^{-1}$ catalyst, 30 μ M ENR | • UV (CEL-LAM500, 50mW $cm^{-2}$, λ < 365 nm) | 99.1 % (10 min) | | |
| | | | | • Simulated solar irradiation (CEL-LAX500, wavelength range: 200-1000 nm, 100 mW $cm^{-2}$) | 72.3 % (10 min) | | |
| | | LEV | 0.1 g $L^{-1}$ catalyst, 30 μ M LEV | • UV (CEL-LAM500, 50 mW $cm^{-2}$, λ < 365 nm) | 98.5 % (10 min) | | |
| | | | | • Simulated solar irradiation (CEL-LAX500, wavelength range: 200-1000 nm, 100mW $cm^{-2}$) | 81.9 % (10 min) | | |
| MnO@ $MnO_x$ | Mesoporous microspheres | LEV | 0.1 g $L^{-1}$ catalyst, 30 μ M LEV | Simulated sunlight irradiation (CEL-LAX500, Beijing China Education Au-light Co., Ltd) | | 98.1 % (30 min) | $•O_2^-$, $^1O_2$, $SO_4^{•-}$ and •OH | [73] |
| $gC_3N_4$/ $Bi_2O_3$@ N-HMCs | Hollow mesoporous spheres | CFH | 0.1 g $L^{-1}$ catalyst, 10 mg $L^{-1}$ CFH | Visible-light irradiation (300 W Xe lamp, λ > 420 nm) | | 78.06 % (60 min) | •OH, $•O_2^-$ and $h^+$ | [152] |
| $MoS_2$ /C | Hollow nanospheres | LEV | 0.1 g $L^{-1}$ catalyst, 0.9 g $L^{-1}$ PMS, 30 μ M LEV | Visible-light irradiation (300 W Xe lamp, λ > 420 nm) | | 97 % (160 min) | $SO_4^{•-}$ and •OH | [166] |

### *3.4. Sulfonamides antibiotics*

Sulfonamides (SDs) are generally a white or light yellow crystalline powder, odorless, and easy to deteriorate. In terms of solubility, most of these drugs have low solubility in water, and the aqueous solution is usually alkaline and stable. The mechanism of action of SDs is that they can participate in the process of competing for dihydrofolate synthase so that p-aminobenzoic acid cannot participate in the synthesis of dihydrofolate. Dihydrofolate is an important component of the corresponding bacterial growth, which inevitably inhibits the growth and reproduction of bacteria [167]. SDs are now also widely used in animal breeding and become important drugs for the prevention of corresponding diseases [168, 169]. In addition, the application history of SDs in human clinical medicine is relatively long. If SDs are frequently used, some of the more sensitive strains will develop resistance due to the long-term and repeated use of such drugs. After resistance to a sulfonamide drug, due to the similar structure of sulfonamide drugs, the drug-resistant bacteria will also gradually develop resistance to other sulfonamides, and it is difficult to exert antibacterial effects when these drugs are used [170]. Similarly, drug residues in animal foods will enter the human body through the food chain, and the human body will indirectly develop drug resistance, which will bring certain difficulties to the treatment of corresponding diseases [171].

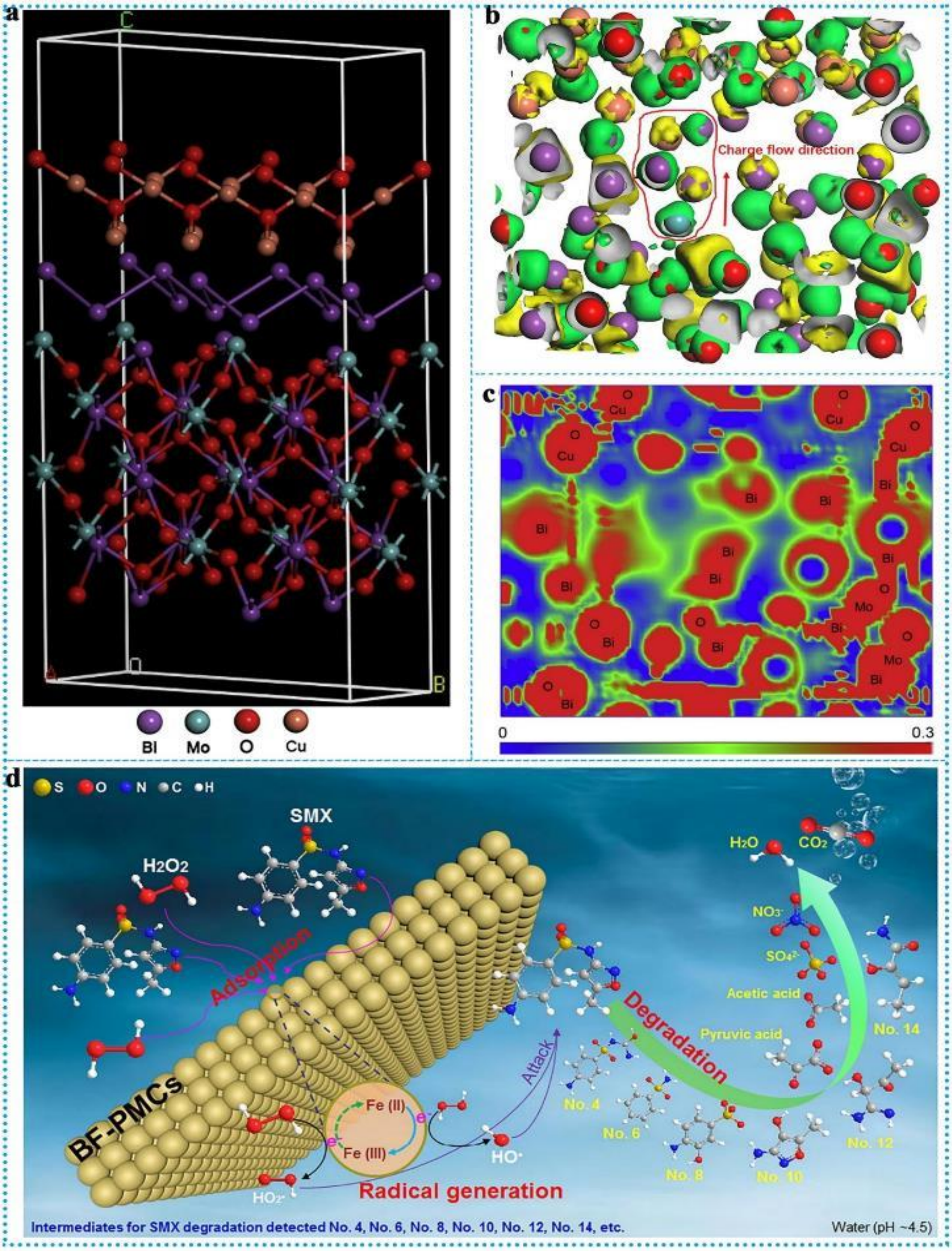

**Fig. 21.** (a) Crystal structures of $Cu_2O/Bi/Bi_2MoO_6$. (b) Charge difference distribution with charge accumulation in yellow and depletion in green, and (c) Electronic location function of $Cu_2O/Bi/Bi_2MoO_6$. Reprinted with permission from Ref. [172]. (d) Possible

degradation mechanism of SMX in heterogeneous BF-PMCs-driven Fenton-like system. Reprinted with permission from Ref. [173].

Xu et al. [172] synthesized $Cu_2O/Bi/Bi_2MoO_6$ by wet impregnation, in which Bi spheres and $Cu_2O$ particles were immobilized in hollow microspheres on the surface of $Bi_2MoO_6$. This ternary plasma Z-type heterojunction exhibited excellent photocatalytic activity. Theoretical calculations confirmed the interactions and e-properties among $Cu_2O$, Bi, and $Bi_2MoO_6$. **Fig. 21a-b** shows that the underlying Bi atoms on the metallic Bi surface acquire light-induced e- from Mo and O atoms on $Bi_2MoO_6$ surfaces, and then its upper Bi atoms transfer the e- to the Cu atoms on the $Cu_2O$ surface. **Fig. 21c** shows a strong covalent interaction between Mo, Bi, and O atoms on the $Bi_2MoO_6$ surface, Cu and O atoms on the $Cu_2O$ surface, Bi atoms on the Bi surface, and further indicating that the charge carriers are mainly generated on $Cu_2O$, Bi, and $Bi_2MoO_6$ surfaces, which in turn forming charge transfer channels between $Bi_2MoO_6$ and $Cu_2O$. In addition, Hu et al. [173] also reported the use of nanostructured BF-PMCs to drive heterogeneous Fenton-like oxidation for efficient degradation of sulfamethoxazole (SMX). The possible degradation pathways of some key byproducts (i.e., $C_4H_4NO_2$, $C_8H_{11}N_3O_4S$, etc.) were thoroughly analyzed in depth (**Fig. 21d**).

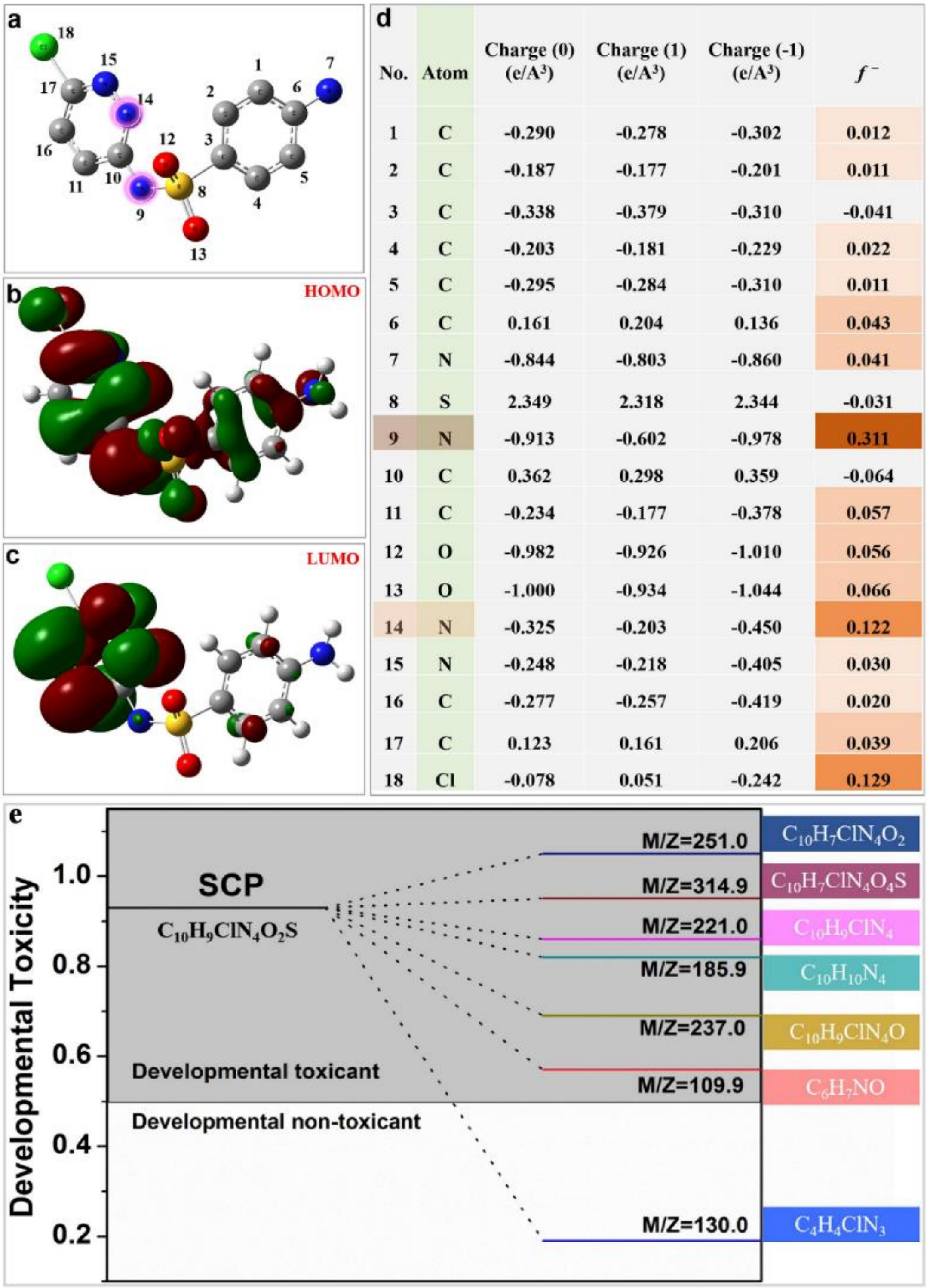

**Fig. 22.** (a) Chemical structure, (b) HOMO, (c) LUMO, and (d) NPA charges and Fukui index ($f^-$) of SCP. (e) The developmental toxicity of the intermediates produced by the degradation of SCP was evaluated by QSAR prediction using the T.E.S.T. Reprinted with permission from Ref. [174].

Tao et al. [174] found a simple way for the in-situ construction of sea urchin-like $WO_3$ modified by $Co(OH)_2$ nanoparticles ($Co(OH)_2/WO_3$, 10 wt%), which could effectively remove sulfachlorpyridazine (SCP, 100% degradation efficiency) by PMS activation. Herein, the adsorption process ($Co^{2+} \rightarrow CoOH^+$) of sea urchin-like $WO_3$ surface is the rate-limiting step of PMS activation and free radical generation. **Fig. 22** explores the relationship between the byproducts of SCP molecules and free radicals attacking. **Fig. 22a** shows the chemical structure of SCP, in which $SO_4^{\bullet-}$ and •OH radicals are considered electrophilic radicals and tend to attack the active sites that are prone to losing $e^-$. The highest occupied molecular orbital (HOMO) of SCP (**Fig. 22b**) shows that $e^-$ escapes easily because it is easily attacked by $SO_4^{\bullet-}$ and •OH free radicals. The corresponding lowest unoccupied molecular orbital (LUMO) is shown in **Fig. 22c**. Then, the natural population analysis (NPA) charges distribution and Fukui index ($f^-$) of SCP atoms calculated based on DFT calculation are listed in **Fig. 22d**, indicating that the higher $f^-$ the value of the site on SCP atoms was more easily to lose $e^-$ and attacked by free radicals. Therefore, the degradation pathway of SCP was explored in $Co(OH)_2/WO_3$. In pathway 1, the S-N in the sulfonamide group of SCP was attacked by $SO_4^{\bullet-}$ and •OH. Further N-N bond breakage might be oxidized. In pathway 2, the aniline radical cation and the cation of SCP undergo intermolecular Smile-type rearrangement and $SO_2$ extrusion to generate products. In pathway 3, SCP was directly oxidized through the hydroxylation of the aniline ring. Therefore, the conclusion drawn was that SCP had a similar SDs degradation pathway as previously reported (i.e. aniline, sulfonamide, and N-containing heterocycles). Furthermore, intermediates for SCP degradation were then evaluated by Quantitative Structure-Activity Relationship (QSAR) prediction using the Toxicity Estimation Software Tool (T.E.S.T.) (**Fig. 22e**). Besides, **Table 8** shows the photocatalytic degradation of SDs by different SMOs. It is clear to see the advantages of different SMOs play a crucial role in the photocatalytic degradation of SDs.

**Table 8** Application of SMOs in the degradation of SD

| Catalysts | Morphology | Target pollutant | Reaction conditions | Performance | Mechanism | Refs. |
|---|---|---|---|---|---|---|
| $Cu_2O/Bi/Bi_2MoO_6$ | Hollow spheres | SDZ | [SDZ] = 10 mg $L^{-1}$, a certain amount of catalyst, visible light (λ > 420 nm) | 98.6% (100 min) | •OH, $•O_2^-$ and $h^+$ | [172] |
| $Fe_3O_4@MoS_2$ | Spheres | SA | [PMS] = 1.0 mM, [SA] = 20 mg $L^{-1}$, [catalyst]= 0.4 g $L^{-1}$ | 99.8 % (15 min) | $SO_4^{•-}$ and $^1O_2$ | [175] |
| $Co(OH)_2/WO_3$ | Urchin-like spheres | SCP | [SCP] = 0.01 mM, [PMS] = 0.05 mM, [catalyst] = 0.1 g $L^{-1}$, pH = 7.0 ± 0.2 | 100% (2 min) | $SO_4^{•-}$ and •OH | [174] |
| $BiFeO_3/Bi_2Fe_4O_9$ | Plate-like shape assembled with spheres | SMX | [SMX] = 1.5 mg $L^{-1}$, [catalyst] = 0.2 g $L^{-1}$, [$H_2O_2$] = 70 mM and pH = 4.5 | 95% (270 min) | •OH | [173] |
| $NH_2$-UiO-66 | Spheres | STZ | [STZ] = 10 mg $L^{-1}$, [catalyst] = 0.2 g $L^{-1}$, light (300W, intensity =100 mW•$cm^{-2}$ and λ> 400 nm) | 100% (60 min) | •OH, $•O_2^-$ and $h^+$ | [176] |
| $Fe_3S_4$ | Spheres | SDZ | [SMX] = 5 mg $L^{-1}$, [catalyst] = 0.3 g $L^{-1}$, [$H_2O_2$] = 2 mM, pH = 5, T = 25 °C, visible light irradiation | 100 % (40 min) | •OH, $•O_2^-$ and $^1O_2$ | [177] |
| Fe-Mn-Cu | Hollow spheres | SMX | [SMX] = 5 mg $L^{-1}$, [catalyst] = 0.2 g $L^{-1}$, [PMS] = 0.4 mM, pH = 7, T = 25 °C | 100 % (60 min) | $SO_4^{•-}$, •OH and $^1O_2$ | [178] |

### *3.5. Other antibiotics*

In addition to the antibiotics described above, other antibiotics such as fluoxetine (FLX) [179], carbamazepine (CBZ) [180], metronidazole (MNZ) [181], and 2-Mercaptobenzothiazole (MBT) [182] have been concerned owing to their nondegradable, accumulation, toxicity and potential carcinogenicity. These antibiotics are briefly described as follows: (1) FLX is a pharmaceutical compound that is widely used to address anxiety and depression; (2) CBZ, an anticonvulsant drug commonly used to heal depression and epilepsy, is ubiquitous in the environment; (3) MNZ, an antibiotic that targets the treatment of diseases caused by anaerobic microorganisms (e.g., protozoa and bacteria), has been detected in drinking water, surface water and municipal wastewater in many areas; (4) MBT, an antibiotic intermediate with a long half-life, is commonly used in the synthesis of pesticides, cephalosporins and as a metal detection reagent, is abundant in the environment. Considering the harm of these antibiotics to human health and the ecosystem, it is necessary to take effective treatment ways to degrade the aqueous solution of these antibiotics. **Table 9** shows the application of SMOs in degradation of other antibiotics including BPA, acetaminophen (AP), FLX, CBZ, MNZ, and MBT.

**Table 9** Application of SMOs in the degradation of other antibiotics.

| Catalysts | Morphology | Target pollutant | Reaction conditions | Performance | Mechanism | Refs. |
|---|---|---|---|---|---|---|
| N-doped $TiO_2$ | Hollow sphere | BPA | • [catalyst] = 0.5 g $L^{-1}$, [BPA] = 5 mg $L^{-1}$, pH = 10.0<br>• LED strip (10 mm wide, 2 mm thick and emission wavelength λ = 465 nm) | 100% (360 min) | •OH | [183] |
| 3U-BiO | Flower-like sphere | BPA | • [catalyst] = 0.8 g $L^{-1}$, [BPA] = 10 mg $L^{-1}$, pH = 6.8<br>• 350 W mercury lamp (UV lamp wavelength of 365 nm) | 90% (30 min) | $h^+$, •OH and $•O_2^-$ | [184] |
| $TiO_2$ | Hollow mesoporous microspheres | AP | • [catalyst] = 0.1 g $L^{-1}$, [AP] = 50 mg $L^{-1}$<br>• 500 W mercury lamp irradiation | 94% (60 min) | \ | [185] |
| $MoS_2/CuFe_2O_4$ | Nanospheres | FLX | • [catalyst] = 0.1 g $L^{-1}$, [PMS] = 1 mM, [FLX] = 20 mg $L^{-1}$, pH = 6.9 | 97.7% (20 min) | $SO_4^{•-}$ and •OH | [186] |
| $Fe_3O_4$/BiOBr @HC | Microsphere | CBZ | • [catalyst] = 0.6 g $L^{-1}$, [CBZ] = 10 mg $L^{-1}$<br>• The 50 W visible LED lamp (dominant wavelength at 475 nm, and luminous flux at 3285.16 lm) | 100% (30 min) | $h^+$, •OH and $•O_2^-$ | [187] |
| Pt@$CeO_2$ @$MoS_2$ | Hollow sphere | CBZ | • 20 mg catalyst, [CBZ] = 2 mg $L^{-1}$, [PMS] = 0.76 mM, pH = 4<br>• Light source: 300 W Xe lamp with 420 nm cut filter | 100% (30 min) | •OH, $•O_2^-$, $SO_4^{•-}$ and $^1O_2$ | [188] |
| CoMgFe-LDO | Spheres | CBZ | • [catalyst] = 0.02 g $L^{-1}$, [CBZ]= 5 mg $L^{-1}$, [PMS] = 0.2 mM, pH = 5.8 | 73.4% (20 min) | $SO_4^{•-}$ and •OH | [189] |
| ACO | Spheres | MNZ | • [catalyst] = 0.5 g $L^{-1}$, [MNZ]= 10 mg $L^{-1}$, [PMS]= 1.5 mM, pH = 7 | 100% (40 min) | $SO_4^{•-}$ | [190] |
| $CoAl_2O_4$@AP | Spheres | MNZ | • [catalyst] = 20 g $L^{-1}$, [MNZ] = 20 mg $L^{-1}$, [PMS]=1 mM, pH = 6.48 | 97% (100 min) | $SO_4^{•-}$ and $^1O_2$ | [191] |
| $SnO_2$ NPs | Spheres | CBZ | • [catalyst] = 0.5 g $L^{-1}$, [CBZ]= 5 mg $L^{-1}$, pH = 5<br>• 36 W UV-C light | 97% (60 min) | •OH, $h^+$ and $•O_2^-$ | [192] |
| Co-ZnO | Microspheres | MBT | • [catalyst] = 0.5 g $L^{-1}$, [MBT] = 20 mg $L^{-1}$<br>• visible-light irradiation (300 W Xe lamp, λ > 420 nm) | 97.3% (50 min) | $h^+$, $•O_2^-$and $e^-$ | [193] |
| $Bi_2WO_6$/ $In(OH)_3$ | Flower-like sphere | MBT | • [catalyst] = 0.5 g $L^{-1}$, [MBT] = 10 mg $L^{-1}$ • 250 W Xe lamp with a cutoff filter (λ > 420 nm) | 93.1% (120 min) | $h^+$ and $•O_2^-$ | [182] |

Srinivasan et al. [183] developed hollow sphere N-doped $TiO_2$ promises (NhT) by templating PS and ammonia treatment. Compared to the $TiO_2$ powder (PT), NhT exhibited several advantages including high specific surface area, high porosity, and low packing density. These advantages will be beneficial for improving photon efficiency and dispersion in the photoreactor. As expected, NhT could remove 90% of BPA within 2 h under the blue LED irradiation. The removal efficiency of NhT for BPA was almost three times that of PT. Moreover, Liou et al. [185] designed a template-free solvothermal method to prepare hollow core-shell mesoporous $TiO_2$ microspheres with an average diameter of 1.2 μm and shell thickness of 50 nm for efficient photocatalytic degradation of AP. This hollow sphere could improve light capture efficiency and high surface volume ratio. Meanwhile, the mesoporous channels on the shell facilitated the free diffusion of AP into the hollow interior, ensuring full contact between AP and ROS for enhancing photocatalytic activity. Kumar et al. [194] reported a method for synthesizing $Fe_3O_4$-$BiVO_4$/$Cr_2V_4O_{13}$ (FBC) heterojunction via an ultrasonication and hydrothermal route. $Fe_3O_4$ can not only mediate the formation of Z-type nano-heterojunction $BiVO_4$/$Cr_2V_4O_{13}$ with strongly coupled but also imparts certain magnetic properties to the catalyst to facilitate recycling. The novel FBC with high interfacial transfer can degrade 99.2% of FLX after 60 min reaction under visible light. Finally, the redox mediators facilitated between $Fe_3O_4$ and $Cr^{6+}$/$Cr^{2+}$ were proposed. It is very necessary to study the intermediate products of antibiotics because secondary pollution is formed after intermediate products are discharged into the water body. Therefore, Bai et al. [186] studied the byproducts of FL degradation in the $MoS_2/CuFe_2O_4$/PMS system, in which the degradation pathways for the degradation of FLX were proposed by analyzing the structures of intermediates.

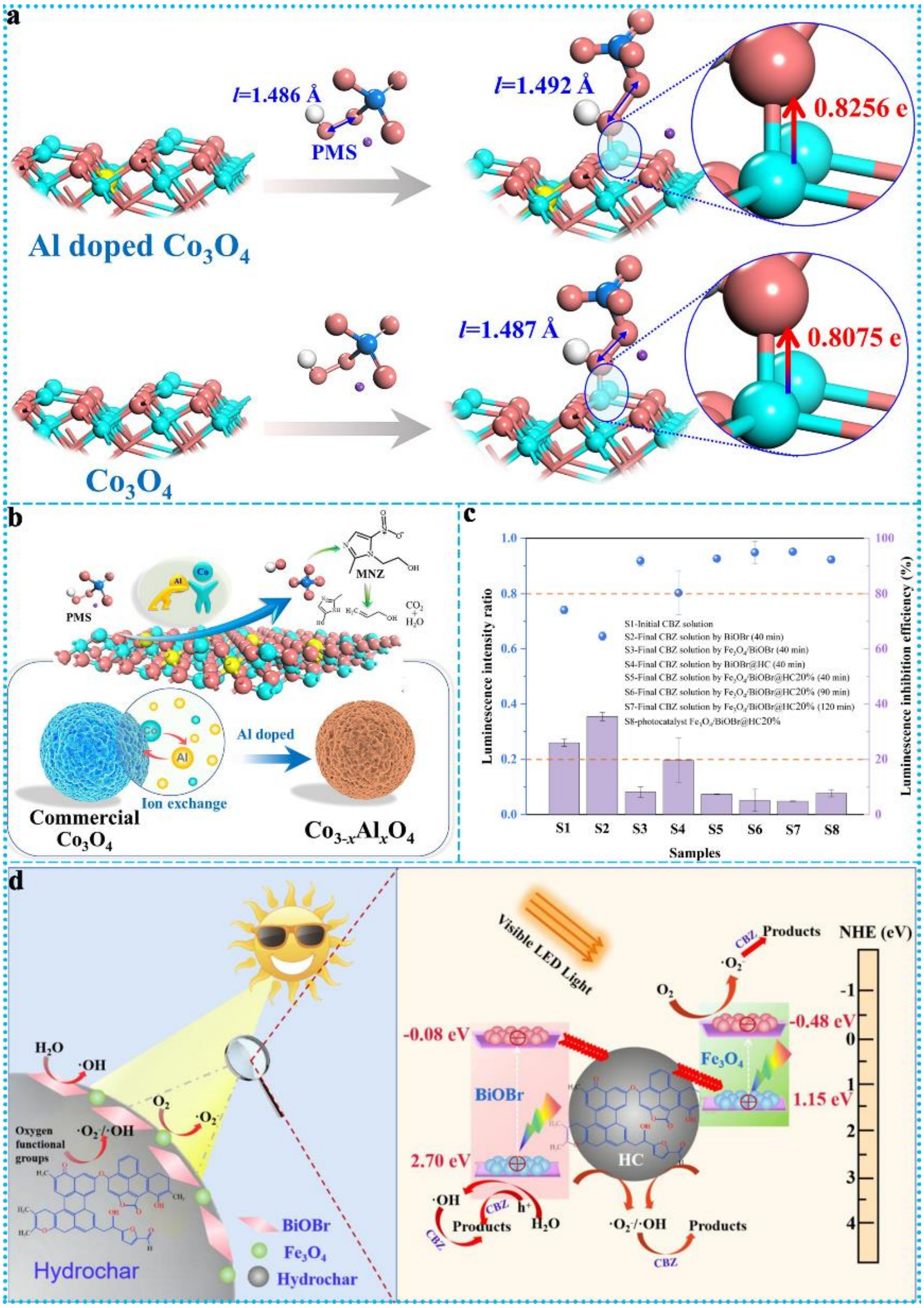

**Fig. 23.** Fenton and Fenton-like degradation of MNZ: (a) DFT configurations of Al-doped $Co_3O_4$ and blank $Co_3O_4$. (b) The corresponding catalytic mechanism in the ACO-

2/PMS system for degradation of MNZ. Reprinted with permission from Ref. [190]. Photocatalytic degradation towards CBZ by $Fe_3O_4$/BiOBr@HC20%: (c) Toxicity assessment in terms of luminescence intensity of Vibrio Fischer and (d) Proposed photocatalytic mechanism. Reprinted with permission from Ref. [187].

Li et al. [190] reported the formation of Al-doped $Co_3O_4$ catalysts (ACO) by introducing $Al^{3+}$ ions with catalytically inert into commercial $Co_3O_4$ lattice. Impressively, the ACO has a strong weakening ability to the bond of adsorbed PMS. According to **Fig. 23a**, the PMS adsorbed on ACO has obvious $l_{O\text{-}O}$ bond extension after adsorbing in ACO. The length of the $l_{O\text{-}O}$ bond (1.492 Å) is larger than the $l_{O\text{-}O}$ bond length (1.487 Å) on the PMS adsorbed on blank $Co_3O_4$, suggesting that the existence of Al in ACO could not only enhance the adsorption capacity of PMS and electron transfer capacity during the reaction but also promote the fracture of $l_{O\text{-}O}$ bond of adsorbed PMS. In the ACO-2/PMS system, MNZ can be effectively degraded since doping of Al helps to accelerate the $Co^{2+}/Co^{3+}$ cycle (**Fig. 23b**). Li et al. [187] used an improved hydrolysis method to construct a composite photocatalyst $Fe_3O_4$/BiOBr@HC, which showed rapid surface catalysis and efficient electron transfer ability in the reaction. Through studying the photoabsorption, electron transfer ability, and reactive free radicals, the possible photocatalytic mechanism of CBZ degradation in the $Fe_3O_4$/BiOBr@HC system was proposed, and the low toxicity and high safety of the photocatalytic system were further verified by studying the toxicity test of Vibrio Fischeri, as shown in **Fig. 23c**. The results showed that the intermediates and $Fe_3O_4$/BiOBr@HC itself had no obvious inhibition effect on luminescence. Meanwhile, under irradiation of the visible-LED-light for a reaction of 40 min, the TOC reached 67.41% in the $Fe_3O_4$/BiOBr@HC20% system, demonstrating excellent photocatalytic activity compared to other photocatalysts. **Fig. 23d** shows the possible photocatalytic mechanism of CBZ degradation in the $Fe_3O_4$/BiOBr@HC system. Besides, Hong et al. [189] prepared a heterogeneous CoMgFe-LDO catalyst as an activator for PMS activation to degrade CBZ, and proposed the possible reaction mechanism of the CoMgFe-LDO/PMS/CBZ system, suggesting that the catalytic activity of CoMgFe-LDO could be connected with the positive synergy effect between Co and Fe.

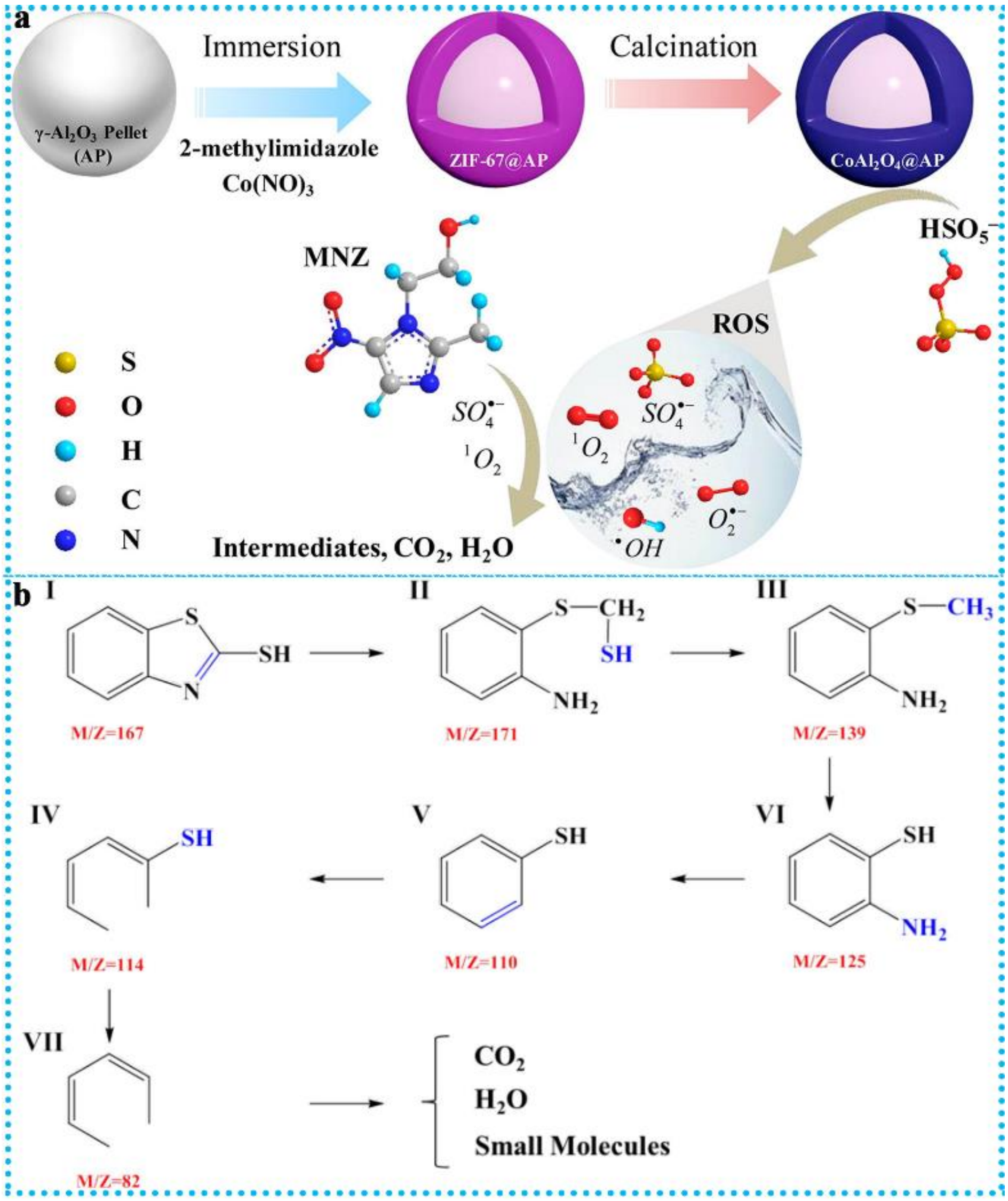

**Fig. 24.** (a) $CoAl_2O_4$@AP/PMS system for degradation of MNZ. Reprinted with

permission from Ref. [191]. (b) Possible intermediate products of degrading MBT over 9-$Bi_2WO_6/In(OH)_3$ composite. Reprinted with permission from Ref. [182].

Meanwhile, Zhu et al. [191] also synthesized a heterogeneous catalyst ($CoAl_2O_4$@AP) for efficient activation of PMS to generate ROS for the degradation of MNZ (**Fig. 24a**). When $CoAl_2O_4$@AP (20 g $L^{-1}$) and PMS (1.0 mM) were added to 20 mg $L^{-1}$ of MNZ, ~97% of MNZ can be removed within 100 min reaction. Huo et al. [182] prepared a novel composite photocatalyst (9-$Bi_2WO_6/In(OH)_3$) through a facile calcination-hydrothermal synthesis method. DFT calculation revealed that in the $Bi_2WO_6$ nanocubes (010) and (001), the energy level difference between the conduction bands and valence bands could build the surface heterojunction, which is conducive to promoting the separation of electron-hole in the reaction process. Therefore, under visible light, 9-$Bi_2WO_6/In(OH)_3$ showed excellent photocatalytic activity for degradation of MBT. The proposed degradations pathways of MBT are shown as **Fig. 24b**. Through addition reaction, the I (m/z = 167) becomes II (m/z = 171), then forms III (m/z = 139) through dropping of −SH, and eventually gets to IV (m/z = 125) and V (m/z = 110) by $-CH_3$ and $-NH_2$ losing, respectively. Through wiping off the −SH and additive reaction, the VI and VII are generated. However, the MBT will eventually break down into small molecules as the reaction proceeds.

In addition to the size effect of SMO-based photocatalysis discussed above, similar phenomena could also be observed in SMO-based Fenton-ike systems. Wang et al. [195] used a one-step hydration method for synthesizing porous $MnO_2$ spheres of different sizes by adjusting the concentration of $KMnO_4$. They also found that the structure-activity relationship of the size effect of $MnO_2$ spheres existed in Fenton-like systems. Specifically, the small-sized $MnO_2$ (PM-S), the medium-sized $MnO_2$ (PM-M) and the large-sized $MnO_2$ (PM-L) were mainly spherical with particle sizes of about 200, 350, and 500 nm, respectively. The complete degradation time of phenol by PM-S (30 min) was shorter than that of PM-M (45 min) and PM-L (60 min) in Fenton-ike systems, which was correlated to the small size, clear hierarchical pores and high surface area of PM-S. Besides, Xie et al. [193] constructed a flower-shaped Co-doped ZnO microspheres heterogeneous photocatalyst and systematically studied its crystal structure, light absorption, and electrochemical performance. The results show that through Co doping, the response of the catalyst to visible light is significantly enhanced with the band gap width decreasing from 3.20 to 2.54 eV. Moreover, the conversion of $Co^{2+}$ to $Co^{3+}$ generates more oxygen vacancies, the photogenerated electrons are transferred to $Co^{3+}$ during the reaction, which effectively inhibits the photogenerated electron-hole recombination and thus improves the photocatalytic activity. Therefore, the 7% Co-ZnO photocatalyst showed great photocatalytic activity, and ~97.3% MBT was removed in 50 min with the rate constant of 0.1498 $min^{-1}$.

## 4. Conclusion and prospect

SMO-based catalysts have widely attracted attention in the photocatalytic degradation of antibiotics due to their unique characteristics including wide band gap, high mobility, and controllable doping capabilities. Many studies have focused on the construction of SMOs with diverse structures (core-shell, porous shell, hollow and yolk-shell, etc.) owing to their highly exposed active sites, excellent catalytic performance, high specific area, large pore volume, and chemical tenability. In this review, the controllable design of SMOs catalysts, their application and mechanisms for the photocatalytic degradation of antibiotics in the environment have been summarized and reviewed, focusing on the various synthetic strategies, the relationship between catalytic activity and different morphologies of SMOs. Hence, we summarize and propose the following perspectives: (1) The hard template method is better applied to construct SMOs with hollow structures, which improves the utilization of the light by realizing multiple scattering and reflection of light. (2) The soft template method is widely used to obtain SMOs with porous structures including microspheres/nanosphere and microflower spheres, which can enhance the specific surface area for exposing more adsorption and reaction active sites and enhancing the mass transfer efficiency of ROS and antibiotics. (3) The template-free method is suited to simplify the synthesis route for the SMOs with multi-shelled hollow structures, which can expose more catalytic active sites for catalytic degradation reaction and reduce the resistance of $e^-$ and $h^+$ to the target pollutants for enhancing the mass transfer efficiency of the reaction system. (4) The template-free method is mainly composed of three main approaches such as Kirkendall effect, Ostwald ripening, and Oriented attachment. The synthesis of a multi-shell hollow structure of SMOs by the template-free method can not only protect the aggregation of internal nanospheres but also allow the effective diffusion of ROS. Finally, the relationship between the main photocatalytic mechanism and morphologies structure of SMOs is obtained by reviewing SMOs (porous spheres, micro flowers, yolk-double shell sphere, yolk-shell sphere, hollow multi-shelled sphere, and hollow spheres) for the degradation of antibiotics (TCs, FQs, SDs, and other antibiotics).

Although great progress has been made in the study of SMOs semiconductor photocatalysts for the degradation of antibiotics, future research still needs to concentrate on the following aspects: (1) The antibiotics degradation efficiency is high in the SMOs photocatalytic system, while the mineralization is slightly low. Therefore, in actual wastewater treatment, combining with other auxiliary treatments including $SO_4^{\bullet -}$-based AOPs, ozone oxidation, electrocatalysis technology, and conventional biological oxidation would need more efforts to largely enhance the mineralization rate. (2) There is still large space to improve the catalytic stability and photo corrosion resistance of SMOs, constructing SMOs-based semiconductor heterojunction photocatalysts (e.g., type I, type II, Z-scheme, and S-scheme) can effectively inhibit the photogenerated electron-hole recombination, enhance the redox properties and broaden the light absorption range of the material, thereby improving the photocatalytic performance. (3) At present, most of the studies are conducted in the simulated wastewater. In addition to the target contaminants, a variety of complex background components, such as heavy metals, inorganic salts, organic matter, and pathogenic microorganisms, will affect the catalytic activity and stability of SMOs in the real wastewater environment. Therefore, significant efforts are still needed to assess the use of these technologies in actual wastewater. (4) To deeply understand the photocatalytic mechanism of SMOs, the applications of DFT, in situ characterization technology, and machine learning in understanding and optimizing SMOs-based systems still require more effort. (5) Photocatalytic degradation of antibiotics using solar energy has extensive application in energy and environmental development. Although target antibiotics have been degraded by SMOs, many degraded intermediates exist in the water due to their weak mineralization ability. What's more, these intermediates may be more toxic than the original antibiotics. Thus, the toxicity of the whole process including intermediates should be considered. In general, this review will be expected to provide the rational design of highly efficient SMOs catalysts for the sustainably photocatalytic removal of different pollutants.

**Data availability**

Data will be made available on request.

**Declaration of Competing Interest**

The authors declare that they have no known competing financial interests or personal relationships that could have appeared to influence the work reported in this paper.

**Acknowledgments**

This work was supported by National Natural Science Foundation of China (22078374), Key-Area Research and Development Program of Guangdong Province (2019B110209003), and the Scientific and Technological Planning Project of

Guangzhou, China (202206010145).